%% file: main.tex
\documentclass{article}
\usepackage[preprint]{neurips_2021}

\input{prelude.tex}
\title{Online Matching in Sparse Random Graphs: Non-Asymptotic Performances of Greedy Algorithm}
\author{%
  Nathan Noiry\\
  Télécom Paris, Palaiseau, France\\
  \texttt{noirynathan@gmail.com }\\
\And
    Flore Sentenac\\
  CREST, ENSAE Paris, Palaiseau, France\\
  \texttt{flore.sentenac@ensae.fr} 
\And 
  Vianney Perchet\\
  CREST, ENSAE Paris, Palaiseau, France\\
  CRITEO AI Lab, Paris, France\\
  \texttt{vianney.perchet@normalesup.org} \\
  }
\begin{document}
\maketitle

\begin{abstract}%
Motivated by sequential budgeted allocation problems, we investigate  online matching problems where connections between vertices are not i.i.d., but they have fixed degree distributions -- the so-called configuration model. We estimate the competitive ratio of the simplest algorithm, \greedy, by approximating some relevant stochastic discrete processes by their continuous counterparts, that are solutions of an explicit system of partial differential equations. This technique gives precise bounds on the estimation  errors,  with arbitrarily high probability as the problem size increases. In particular, it allows the formal comparison between different configuration models. We also prove that, quite surprisingly,  \greedy can have  better performance guarantees than \ranking, another celebrated algorithm for online matching that usually outperforms the former.

\vspace{0.5cm}



\end{abstract}


\section{Introduction}

Finding matchings in bipartite graphs $(\mathcal{U}\cup\mathcal{V},\mathcal{E})$, where $\mathcal{E} \subset \mathcal{U} \times \mathcal{V}$ is a set of edges, is a long-standing problem  with different motivations and approaches \citep{Godsil,Lenka,Lovasz,Salez}. If $\mathcal{U}$ is seen as a set of resources and $\mathcal{V}$ as demands, the objective is  to allocate as many resources  to demands (an allocation - or a matching - between $u$ and $v$ is admissible if $(u,v) \in \mathcal{E}$) with the constraint that a resource is allocated to only one demand and vice-versa. 

Motivated  particularly by practical applications of Internal advertising, the \textsl{online} variant of this problem is  receiving increasing attention (we refer to the  excellent survey \citep{Mehta} for more applications, specific settings, results and techniques). In this case, the set of vertices $\cU$ is present at the beginning  and the graph unveils sequentially: vertices $v \in \cV$ are observed sequentially, one after the other, along with the edges they belong to. An online algorithm must decide, right after observing $v_k$ and its associated set of edges $\mathcal{E}_k:=\{(u,v_k) \in \mathcal{E}\}$  to match it to some other vertex $u \in \cU$, at the conditions that $(u,v_k) \in \mathcal{E}_k$ and $u\in \cU$ has not been matched yet. The performance of an online algorithm is evaluated by its {\it competitive ratio}, which is the ratio between the size of the matching it has created and the highest possible matching in hindsight \citep{Feldman}.

This theoretical setting is particularly well suited for online advertising: $\cU$ is the set of campaigns/ads that an advertiser can run and users $v_1,v_2,\ldots,v_T$ arrive sequentially \citep{Mehta,Manshadi}. Some of them are eligible for a large subset of campaigns, others are not (usually based on their attributes/features, such as the geographic localization, the browsing history, or any other relevant information). The objective of an advertiser (in this over-simplified model) is to maximize the number of  displayed ads. In practice, campaigns/ads are not  displayed only once but have a maximal budget of impressions (say, a specific ad can be displayed only 10.000 times each day). A possible trick consists of duplicating the vertices of $\cU$ as many times as the budget. However, this results in strong and undesirable correlations between vertices. It is therefore more  appropriate to consider a  bipartite graphs with \textsl{capacities} and admissible matchings as subsets of edges such that each vertices belong to several different edges, but not more than their associated capacities $\omega \in \mathds{N}$ (a vertex $v\in \cV$ is matched  once while $u \in \cU$ can be matched $\omega_u$ times).

This online matching problem with capacities has been quite extensively studied. It is known that \greedy, which matches all incoming vertices to any available neighbor has a competitive ratio of $1/2$ in the worst case, albeit it achieves $1- 1/e$ as soon as the incoming vertices arrive in Random Order \citep{GoelMehta}. The worst case optimal algorithm is the celebrated \ranking , which achieves $1- 1/e$ on any instance \citep{ranking,rankingprimaldualanalysis,rankingmadesimple}, and also has better guarantees in the Random Order setting \citep{rankingrandomorder}. 

Beyond the adversarial setting, the following stochastic setting has been considered:  there exist a finite set of $L$ ``base'' vertices $v^{(1)},\ldots,v^{(L)}$ associated to base edge-sets $\mathcal{E}^{(1)}, \ldots ,\mathcal{E}^{(L)}$. When a vertex $v_k$ arrives, its type $\theta_k \in \{1,\ldots,L\}$ is drawn iid from some distribution (either known beforehand or not) and then its edge set is set as $\mathcal{E}_k = \mathcal{E}^{(\theta_k)}$. In the context where the distribution is known, algorithms with much better competitive ratio than \greedy\ or \ranking\ were designed \citep{Manshadi,JailletLu,brubach2019online}, specifically with a competitive ratio of $1-2/e^2$ when the expected number of arrival of each type is integral
and $0.706$ without this assumption. Notably, those competitive ratios still hold with Poisson arrival rates rather than a fixed number of arrivals.

On a side note, a vast line of work considers online matching in weighted graphs \citep{asymptoticsadwords,Adwordmehta,Mehta}, which is outside the scope of this paper. However, it is still worth noting that the unweighted graph is a weighted graph with all weights equal.

This model of the stochastic setting is quite interesting but rather strong: it lacks flexibility and cannot be used to represent some challenging instances ( for example when the degrees of each vertex $\cU$ increase linearly with the number of vertices in $\cV$, or when the set $\cU$ of campaigns must be fixed so that the model is well specified, etc...). Another tentative is to consider Erd\H{o}s-R\'enyi graphs assuming  that each possible edge is present in $\mathcal{U} \times \mathcal{V}$ with some fixed probability and independently of the other edges (see \citep{MastinJaillet}). The most interesting and challenging setting corresponds to the so-called {\it sparse} regime where each vertex of $\cU$ has an expected degree independent of the size $n$ of $\cV$, which amounts to take a probability of connection equal to $c/n$. Interestingly enough, even the analysis of the simplest \greedy\ algorithm is quite challenging and already insightful in those models \citep{Borodin,Arnosti,Dyer,MastinJaillet}. Unfortunately, although this Erd\H{o}s-R\'enyi model is compatible with growing sets $\cU$ and $\cV$, it also turns out to be quite restrictive. The main problem is that the approximate Poisson degree distribution of the vertices has light-tail and does not allow for the appearance of the so-called {\it scale-free property} satisfied by many real-world networks \citep{barabasi2000scale, van2016random}.


We therefore consider a more appropriate random graphs generation process called \textsl{configuration model}, introduced by \citep{BenderCanfield78} and \citep{Bollobas80}. The optimal matching of this model has been computed in \citep{Salez}. The configuration model is particular well suited to handle different situations such as the following one. Assume  that campaigns can either be ``intensive'' (with many eligible users) or ``selective/light'' (few eligible users), with an empirical proportion of, say, 20\%/80\%. Then whether an advertiser handle 100 campaigns at the same time or 10.000, it will always have  roughly this proportion of intensive vs.\ light campaigns. Similarly, some users are more valuable than others, and are therefore eligible to more campaigns than the others; the proportion of each type being independent of the total population size. 
The configuration model accommodates these observations by  basically drawing iid degrees for vertices $\cU$ and $\cV$ (accordingly to some different unknown distributions for $\cU$ and $\cV$) and then by finding a graph such that those degrees distribution are satisfied (up to negligible errors); as a consequence, the graphs generated are \textsl{sparse}, in the sense that the number of edges grows linearly with the number of vertices.

We  investigate the performances (in terms of expected competitive ratio) of the greedy matching algorithm  in configuration models  and we provide explicit quantitative results using stochastic approximation techniques \citep{Wormald}; we prove that the increasing size of the random matching created is arbitrarily close to the solution of some explicit ODE. Solving the latter then gives in turn the solution to the original problem.

The remaining of the paper is organized as follows. Section \ref{SE:Model} describes precisely the problem and Theorem \ref{theo:main_theo} is our first main result: it describes the performances of \greedy\ in the capacity-less problem.  The proof of Theorem  \ref{theo:main_theo} is delayed to \cref{Appendix_main_theo}, but the main ideas and intuitions are provided in Section \ref{SE:proof_main}.  The online matching with capacities problem is treated in \cref{ap:generalresult}. 

\section{Online Matching Problems; Models and main result}\label{SE:Model}
Consider a bipartite graph with capacities $G = (\cU,\cV,\cE,\omega)$ where $\cU=\{1,\ldots, N\}$ and $\cV=\{1,\ldots, T\}$ are two finite set of vertices, $\cE \subset \big\{ (u,v), \, u \in \cU, \, v \in \cV \big\}$ is the set of edges and $\omega: \cU \to \mathds{N}_*$ is a capacity function. 
 A matching $M$ on $G$ is a subset of edges $e \in \cE$ such that any vertex $v \in \cV$ is the endpoint of at most one edge $e \in M$ and any vertex $u \in \cU$ is the endpoint of at most $\omega_u$ edges in $M$.  We will denote by $\cM$ the set of matchings on $G$; the optimal matching $M^* \in \cM$ is the one (or any one) with the highest cardinality, denoted by $|M^*|$.
  
The batched matching problem consists in finding any optimal matching $M^*$ given a graph with capacities $G$; the online variant might be a bit more challenging, as the matching is constructed sequentially. Formally, the set of vertices $\cU$ and their capacities $\omega$ are known from the start and vertices $v \in \cV$ arrive sequentially (with the edges they belong to) and $M_0 = \emptyset$. At stage $t \in \mathds{N}$ -- assuming a matching $M_{t-1}$ has been constructed --,  a decision maker observes a new vertex\footnote{Although the order of arrival is irrelevant to the models we studied, it could have an impact on other models.} $v_t$ and its associated set of edges $\{ (u,v_t); u \in \cE\}$. If possible, one of these edges $(u_t,v_t)$ is added to $M_{t-1}$, with the constraint that $M_t = M_{t-1}\cup\{(u_t,v_t)\}$ is still a matching. The objective is to maximize the size of the constructed matching $M_T$. The classical way to evaluate the performances of an algorithm is the \textsl{competitive ratio}, defined as $|M_T|/|M^*| \in [0,1]$ (the higher the better).

\subsection{Structured online matching via Configuration Model}
\label{SE:ConfigModel}
As mentioned before, the online matching problem can be quite difficult without additional structure. We will therefore assume that the vertex degrees in $\cU$ and $\cV$ have (at least asymptotically in $N$ and $T$) some given subGaussian\footnote{Actually, we only need that $\pi_\cU$ and $\pi_\cV$ have some finite moment of order $\gamma >2$.} distributions $\pi_\cU$ and $\pi_\cV$, of respective expectation $\mu_\cU$ and $\mu_\cV$ and respective proxy-variance $\sigma_\cU^2$ and $\sigma_\cV^2$. Those numbers are related in the sense that we  assume\footnote{In the general case,  consider $T= \lfloor N\mu_\cU/ \mu_\cV\rfloor$. The proof is identical, up to a negligible $1/N$ error term} that $T = \frac{\mu_\cU}{\mu_\cV}N \in \mathds{N}$.   Given those degree distributions, the graphs we consider are random draws from a bipartite configuration model described below; for the sake of clarity, we first consider the capacity-less case (when $\omega_u = 1$ for all $u \in \cU$).

\medskip

Given $\pi_\cU$ and $\pi_\cV$ and $N,T \geq 1$, let  $d^\cU_1,\ldots, d^\cU_N \in \N \stackrel{\mathrm{i.i.d.}}{\sim} \pi_\cU$  and
$d_1^\cV,\ldots, d_T^\cV\in \N \stackrel{\mathrm{i.i.d.}}{\sim} \pi_{\cV}$
be independent random variables; intuitively, those numbers are respectively the number of half-edges attached to vertex in $\cU$ and $\cV$. Consider also two extra random variables 
$$
d^{\cV}_{T+1}= \max\big\{ \sum_{i=1}^{N} d_i^{\cU} - \sum_{j=1}^{T} d_j^{\cV} \ , 0 \big\} \quad \text{and}\quad
d^{\cU}_{N+1}= \max\big\{ \sum_{j=1}^{T} d_j^{\cV} - \sum_{i=1}^{N} d_i^{\cU} \ , 0 \big\}$$
so that  equality between total degrees holds, i.e., $\sum_{i=1}^{N+1} d_i^{\cU} = \sum_{j=1}^{T+1} d_j^{\cV}$.
Finally, a random (capacity-less) bipartite graph denoted by $\mathbf{CM}(\mathbf{d}^{U}, \mathbf{d}^{V})$  is constructed  with a uniform pairing of half-edges of\ $\cU \cup \{N+1\}$ with half-edges of $\cV\cup\{T+1\}$ and removing  vertices $T+1$ and $N+1$ and their associated edges. These two artificially added vertices are just here to define a pairing between half-edges. Notice that, by the law of large numbers and since $T = (\mu_\cU / \mu_\cV) N$, $d^\cV_{T+1} = o(N)$ and $d^\cU_{N+1} = o(N)$ almost surely\footnote{And even $\mathcal{O}(\sqrt{N})$ with probability exponentially large in $N$ as both distributions are sub-Gaussian. So the effects of those additional vertices can be neglected.}. 

The bipartite configuration model $\mathbf{CM}(\mathbf{d}^\cU,\mathbf{d}^\cV)$ is then the random graph obtained by a uniform matching between the half-edges of $\cU$ and the half-edges of $\cV$, where the random sequences $\mathbf{d}^{\cU}=(d_i^\cU)_i$ and $\mathbf{d}^{\cV}=(d_j^\cV)_j$ are defined as above.

\subsection{Competitive ratio of \greedy\ algorithm. Main result}

The first question to investigate in this structured setting is the computation of the (expected) competitive ratio of the simple  algorithm  \greedy. It constructs a matching by sequentially adding any admissible edge uniformly at random. Describing it and stating our results require the following additional notations:  for any $e=(u,v) \in E$,  $u(e)=u$ (resp. $v(e)=v$) is the extremity of $e$ in $\cU$ (resp. $\cV$);   the generating series of $\pi_\cU$ and $\pi_\cV$ are denoted by $\phi_\cU$ and $\phi_\cV$ and are defined as
\begin{equation*}\label{eq:definition_generatingseries}
\phi_\cU(s) :=  \sum\limits_{k \geq 0} \pi_\cU(k) s^k \quad \quad \text{and} \quad \quad \phi_\cV(s) :=  \sum\limits_{k \geq 0} \pi_\cV(k) s^k.
\end{equation*}

Our first main theorem, stated below,   identifies the asymptotic size of the matching generated by \greedy\ on the bipartite configuration model we have just defined. As the batched problem (i.e., computing the size of the optimal matching $M^*$) is well understood \citep{Salez}, this quantity is sufficient to derive competitive ratios. Again, for the sake of presentation, we first assume that all capacities are fixed, equal to one; the general case is studied later on.

\begin{theorem}\label{theo:main_theo}\textbf{(Performances of \greedy\  in the capacity-less case)} 

Given $N \geq 1$ and $T= \frac{\mu_\cU}{\mu_\cV} N$, let $\mathrm{M}_T$ be the matching built by \greedy\ on $\mathbf{CM}(\mathbf{d}^{U}, \mathbf{d}^{V})$ then the following  convergence in probability holds:
\begin{equation*}\label{eq:convergence_probability}
\frac{|\mathrm{M}_T|}{N} \overset{ \mathbf{P}}{ \underset{ N \rightarrow +\infty}{ \longrightarrow}} 1 - \phi_\cU(1-G(1)).
\end{equation*}
where $G$ is the unique solution of the following ordinary differential equation:
 \begin{equation}\label{eq:edo_thm_1}
      G'(s) =  \frac{1 - \phi_\cV \left( 1 - \frac{1}{\mu_\cU} \phi_\cU'\left( 1 - G(s) \right) \right)}{\frac{\mu_\cV}{\mu_\cU} \phi_\cU'( 1 - G(s)) } ; \quad G(0)=0.
      \end{equation}
Moreover, for any $s \in [0,1]$, if $M_{T}(s)$ is the matching obtained by \greedy\ after seeing a proportion  $s$ of vertices of $\cV$, then 
\begin{equation}\label{eq:ODE_on_F_atS}  \frac{|\mathrm{M}_T(s)|}{N} \overset{ \mathbf{P}}{ \underset{ N \rightarrow +\infty}{ \longrightarrow}} 1 - \phi_\cU(1-G(s)).\end{equation}
Convergence rates are  explicit;  with probability exponentially large, at least $1- \zeta N\exp(- \xi N^{c/2})$,
$$
\sup_{s \in [0,1]}\Big| \frac{|\mathrm{M}_T(s)|}{N} - \big(1 - \phi_\cU(1-G(s))\big) \Big| \leq \kappa N^{-c},
$$
where $\zeta,\xi,\kappa$ depend only on the (first two) moments of both $\pi_\cV$ and $\pi_\cU$, and $c$ is some universal constant (set arbitrarily as $1/20$ in the proof).
\end{theorem}


Theorem \ref{theo:main_theo} generalizes to the case with capacities, see  Sections \ref{SE:Weights} and \ref{SE:General}. The details of the proof of Theorem \ref{theo:main_theo} are postponed to Appendix \ref{Appendix_main_theo}, but the main ideas are given in the following Section \ref{SE:proof_main}.
\subsection{Examples, Instantiations and Corollaries}

We provide in this section some interesting examples and corollaries that illustrate the powerfulness of Theorem \ref{theo:main_theo}, and how it can be used to compare different situations.
\subsubsection{$d$-regular graphs}

The first typical example of random graphs are `` $d$-regular '', for some $d \in \mathds{N}$, i.e., graphs such that each vertex has an exact degree of $d$ (to avoid trivial examples, we obviously assume $d\geq 2$). 

It is non-trivial to sample a $d$-regular graph at random, yet it is easy to generate a  random graphs $\mathrm{G}_N$ with  the  configuration model described above, with the specific choices of $\pi_\cU=\pi_\cV=\delta_d$, the Dirac mass at $d$. The downside is that $G_N$ is not exactly a $d$-regular bipartite random graph (as some vertices might be connected more than once, i.e., there might exist multiple edges). However, conditioned to be {\it simple}, i.e, without multiple edges and loops, it has the law of a uniform $d$-regular bipartite random graph. Moreover, the probability of being simple is bounded away from 0 \citep{van2016random}; as a consequence, any property holding with probability tending to $1$ for $\mathrm{G}_N$,   holds with probability tending to $1$ for uniform $d$-regular bipartite random graphs.  Finally, we also mention that Hall's Theorem \citep{frieze2016introduction} implies that  $\mathrm{G}_N$ admits a perfect matching, so that $|M^*|=n$. 


Instantiating Equation \eqref{eq:edo_thm_1} to $d$-regular graphs yields that the competitive ratio of \greedy\ converges, with probability 1, to $1-(1-G(1))^d$  where $G$ is the solution of the following ODE \vspace{-0.5em}
\begin{equation}\label{eq:ODEonF1}
\frac{(1-G(s))^{d-1}}{1 - \left( 1 - (1-G(s))^{d-1}  \right)^d } G'(s) = \frac{1}{d}.
\end{equation}
As expected, had we taken $d=1$, then $G(s)=s$ hence the competitive ratio of \greedy\ is 1 (but again, $d=1$-regular graphs are trivial).
More interestingly, if $d=2$, the ODE has a closed form solution: $G(s)=\exp(\frac{s}{2})-1$, so that the competitive ratio of \greedy\ converges to $4\sqrt{e}-(e+3)\simeq 0.877 \gg 1-\frac{1}{e} \simeq 0.632$, where the latter is a standard bound of the competitive ratio of \greedy (for general, non-regular graphs) \citep{Mehta}.
\paragraph{Solving Equation \eqref{eq:ODEonF1}}
In the general case $d \geq 3$, even if Equation \eqref{eq:ODEonF1} does not have a closed form solution, it is still possible to provide some insights. Notice first that the polynomial $P(X) = 1 - ( 1 - (1-X)^{d-1})^d$ admits $n:=d  (d-1)$ roots, among which there is $1$ with multiplicity $d-1$. If $X$ is another root, then \vspace{-0.5em}
\[  \left(1 - (1-X)^{d-1}\right)^d = 1 \, \, \Leftrightarrow \, \, 1 - (1-X)^{d-1} = e^{ \frac{i k \pi}{d}}, \, \, k = 1, \ldots, d-1.  \]
Therefore, 
\[  (1-X)^{d-1} = 1 - e^{ \frac{i k \pi}{d} }, \]
which admits $d-1$ distinct solutions for each $k = 1, \ldots, d-1$. The resulting $n:=(d-1)^2$ distinct complex, denoted $x_1, \ldots, x_n$, are the roots of  $P(X) / (1-X)^{d-1}$, so the ODE reduces to:
\begin{equation}  \label{eq:EDO_y1}
\frac{y'(t)}{ \prod_{1 \leq i \leq n} y(t) - x_i } = \frac{1}{d}.
\end{equation}
Since the following trivially holds:\[  \frac{1}{\prod_{1 \leq i \leq n} (X - x_i)} =  \sum_{ 1 \leq i \leq n} \frac{1}{\prod_{j \neq i} (x_i-x_j) }  \frac{1}{X-x_i} =: \sum_{ 1 \leq i \leq n}   \frac{a_i}{X-x_i}. \]
it is possible to  integrate Equation \eqref{eq:EDO_y1} in $\sum_{1 \leq i \leq n } a_i \log ( y(t) - x_i)  = \frac{s}{d}+c$ to finally get
\[   \prod_{ 1 \leq i \leq n} (y(t) -x_i)^{a_i}  = C \exp(\frac{s}{d}),  \]
and since $y(0) = 0$, it must hold that  $C =\prod_{1 \leq i \leq n} (-x_i)^{a_i}$. As a consequence, $y(1)$  solves:
\[   \prod_{ 1 \leq i \leq n} (y(1) -x_i)^{a_i} =  e^{1/d} \prod_{1 \leq i \leq n} (-x_i)^{a_i}. \]

Unfortunately, even for $d=3$, the solution somehow simplifies but has no closed form; on the other hand, numerical computations indicate that the competitive ratio of \greedy converges to 0.89 when $d=3$ and $N$ tends to infinity. We provide in Figure \ref{FIG:dregular} the numerical solutions of the ODE for $d$-regular graphs (actually, we draw the functions $1-\phi_U(1-G(s))$ that are more relevant) for various values of $d$; the end-point obtained at $s=1$ indicates the relative performance of \greedy. As expected, those functions are point-wise increasing with $d$ (as the problem becomes simpler and simpler for \greedy\, when $d\geq 2$).

\vspace{-0.5em}
\begin{figure}
\centering
\includegraphics[height=0.3\textwidth]{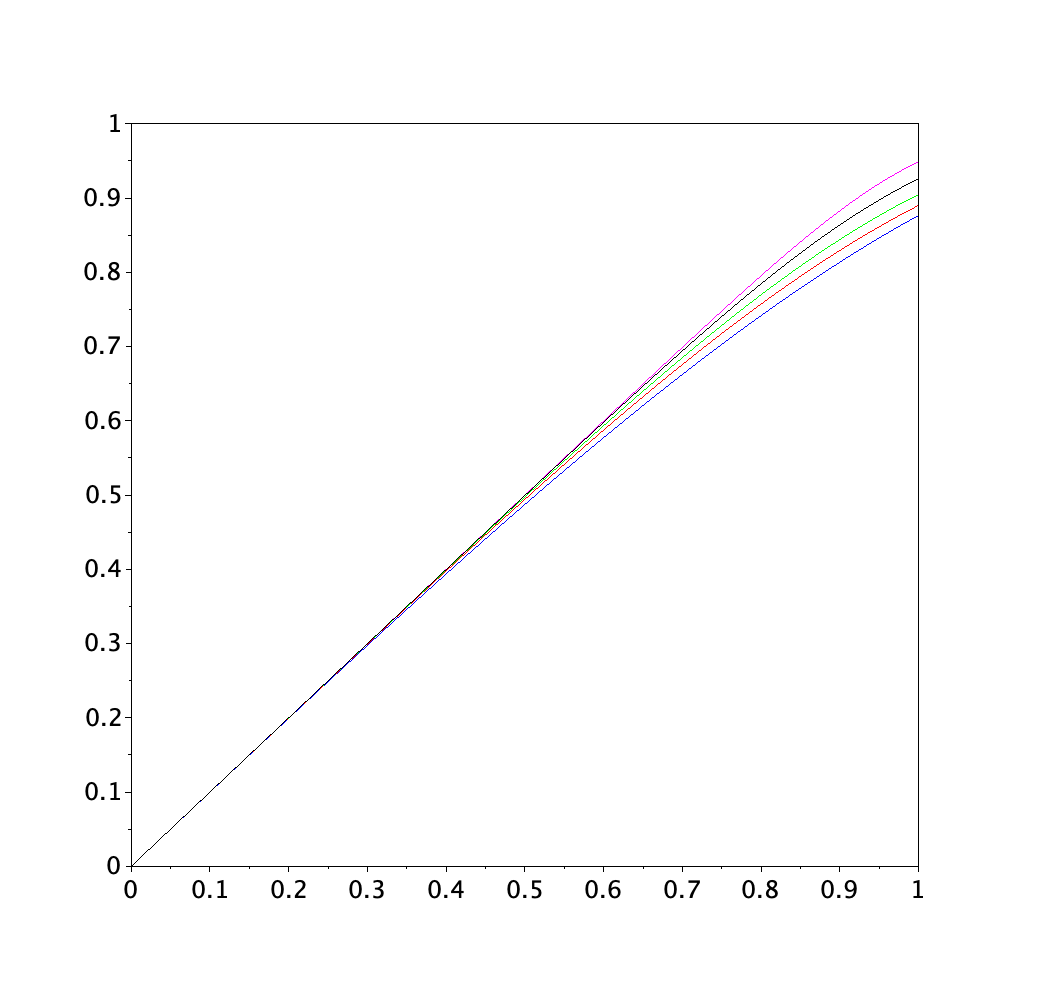}
\includegraphics[height=0.3\textwidth]{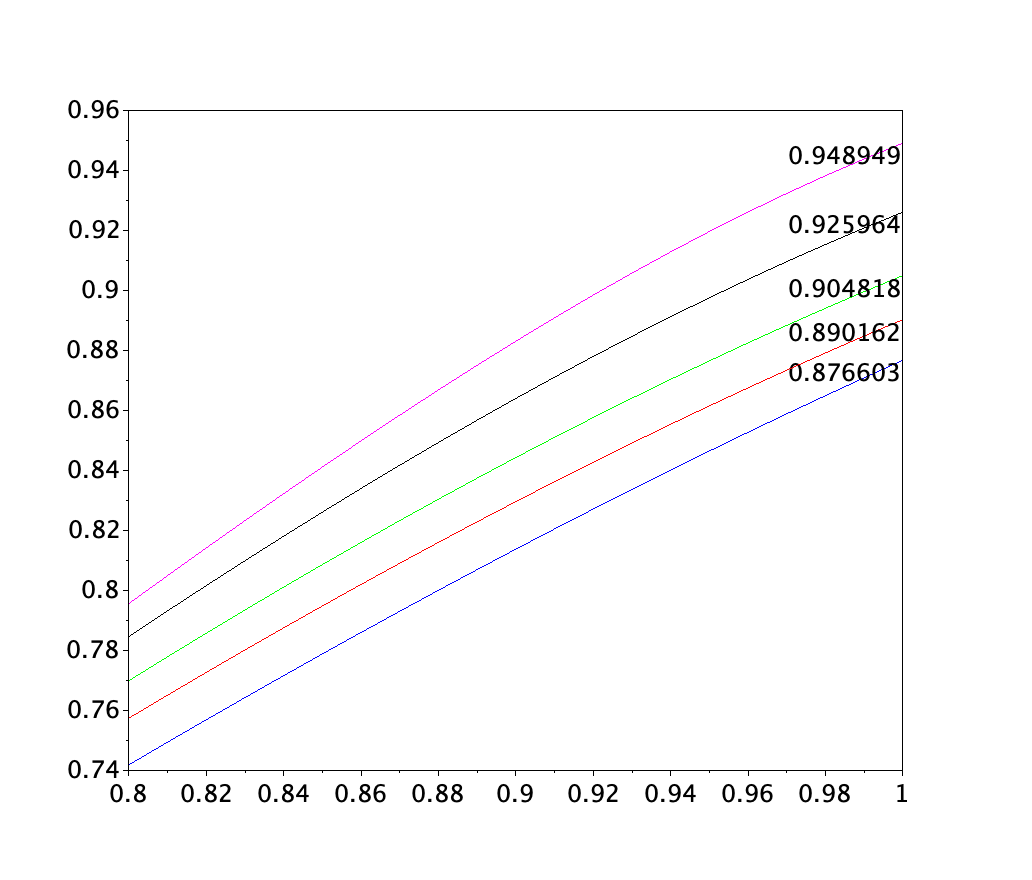}
\caption{Numerical computations (on Scilab, results are almost instantaneous) of \greedy\ performances for $d=2$ (blue), $d=3$ (red), $d=4$ (green), $d=6$ (black) and $d=10$ (magenta). On the left, global solution, on the right, zoom-in on the end points with final values.}
\label{FIG:dregularscore}
\end{figure}
\vspace{-0.5em}


\subsubsection{The Erd\H{o}s-R\'enyi case.} 
In a Erd\H{o}s-R\'enyi graph, there is an edge between two vertices $u \in \cU$ and $v\in\cV$ with some probability $p=\frac{c}{N}$, independently from each others.  As $N$ goes to infinity, the number of edges adjacent to a vertex follows (approximately) a Poisson law of parameter  $c>1$.

As a consequence, we consider the configuration model where $\pi_\cU$ and $\pi_\cV$ are Poisson laws of parameter $c$, which yields $\mu =c$, $\phi_\cU(s) = \e^{c(s-1)}$. In this case, Equation \eqref{eq:edo_thm_1} becomes:
\[  \frac{cG'(s) \e^{-c G(s)}}{1- \e^{ -c \e^{-c G(s)} } } = 1.  \]
The solutions are given by:
\[  G(s) = \frac{1}{c} \log \left(  \frac{c}{ \log( \e^{  k-cs } + 1  ) }  \right),  \]
yielding 
\[  \phi_{X}(1-G(s)) = \frac{1}{c} \log \left(  \e^{k - cs} + 1  \right).  \]

The initial condition $\phi_\cU(1-G(0))=\phi_\cU(1)=1$ gives $e^{k}=e^c-1$, from which we deduce that the number of matches of  \greedy\ is asymptotically proportional to 
\[ 1 - \phi_\cU(1-G(1)) = 1 - \frac{\log \left(  2 - \e^{-c}  \right)}{c},   \]
which recovers, as a sanity check, some existing results  \citep{MastinJaillet}.

\subsubsection{The comparison of different configuration models}

Using Gronwall's Lemma, it is possible to show \cref{theo:main_theo} can be used to compare different configuration models, as in the following Corollary.

\begin{corollary}
Consider two configuration models $\mathbf{CM}_1(\mathbf{d}_1^{U}, \mathbf{d}_1^{V})$ and $\mathbf{CM}_2(\mathbf{d}_2^{U}, \mathbf{d}_2^{V})$, s.t. $\mathbf{d}_1^{U}$ and $\mathbf{d}_2^{U}$ are both drawn i.i.d. from $\pi_U$, $\mathbf{d}_1^{V}$ is drawn i.i.d. from $\pi_V^1$ and $\mathbf{d}_2^{V}$ is drawn i.i.d. from $\pi_V^2$, with $\sum_x x \pi_V^1(x) = \sum_x x \pi_V^2(x)$. If $\phi_V^1(s)\geq \phi_V^2(s)$ for any $s \in (0,1)$, then by denoting respectively $\gamma_1$ and $\gamma_2$  the asymptotic proportion of vertices matched by \greedy\ in  $\mathbf{CM}_1(\mathbf{d}_1^{U}, \mathbf{d}_1^{V})$ and $\mathbf{CM}_2(\mathbf{d}_2^{U}, \mathbf{d}_2^{V})$, it holds that necessarily
$
\gamma_2 \geq \gamma_1.
$
\end{corollary}

For instance, let use assume that the degree distribution on the offline side is fixed. Then the matching size obtained by \greedy\ is asymptotically  larger if  vertices on the online side all have exactly the same degree $d$ rather than if those degrees are drawn from a Poisson distribution with expectation $d$. 

A similar result (with a different criterion) holds with fixed degree distribution on the online side and differing one on the offline side.


\subsection{\greedy\ can outperform \ranking\ !}
Quite surprisingly, we also get that in the configuration model \ranking\ can have a worst competitive ratio than \greedy, which advocates again for its thorough study.
\begin{proposition}\label{prop:greedyvsranking}
On the $2$-regular graph, \greedy\ outperforms \ranking.
\end{proposition}

We conjecture that the above result actually holds for any $d \geq 2$, and more generally for a wide class of distributions $\pi_\cU$ and $\pi_\cV$ (finding a general criterion would be very interesting). The proof of Proposition \ref{prop:greedyvsranking} is provided in \cref{ap:greedyvsranking}. The main idea is that in the $2$-regular graph, \ranking\ is biased towards selecting as matches vertices with two remaining half-edges rather than just one.
Indeed, vertices with only one remaining half edge where not selected previously and thus have a higher rank. The vertices with only one remaining half edge will not get matched in the subsequent iterations, so not picking them as matches is suboptimal. On the other hand, \greedy\ picks any match uniformly at random and does not exhibit such bias.

\section{Ideas of proof of Theorem \ref{theo:main_theo}}
\label{SE:proof_main}
The main idea behind the proof of Theorem \ref{theo:main_theo} (postponed to Section \ref{Appendix_main_theo}) is to show that the random deterministic evolution of the matching size generated by \greedy\ is closely related to the solution of some ODE (this is sometimes called ``the differential equation method'' \citep{Wormald}  or ``stochastic approximations'' \citep{Robbins}. Computing the solution of the ODE is easier - and if not explicitly, at least numerically in intricate cases - than estimating the performances of \greedy\ by Monte-Carlo simulations and it provides qualitative, as well as quantitative, properties. 

Tracking the matching size is actually non-trivial because vertices (in $\cU$ and in $\cV$) have different degrees, hence some of them are more likely to be matched than other. However, in the configuration model, each vertex has the same distribution of degrees before the sequences $\mathbf{d}^\cU$ and $\mathbf{d}^\cV$ are fixed. As a consequence, the proof relies on the three following techniques
\begin{enumerate}
    \item The graph is built sequentially, along with the matching and not beforehand (fixing the "randomness'' at the beginning would be very difficult to handle in the analysis). Thankfully, this does not change the law of the graph generated (this is obviously crucial).
    \item We are not only going to track the size of the matching built as we need to handle different probability of matching (and pairing the graph) for each vertex. As a consequence, we are going to track the numbers of non-matched vertices which have still $i$ half-edges to be paired and the number of already matched-vertices that have $j$ half-edges remaining. This will give one different ODE per value of $i$ or $j$.
     
    Since $\pi_\cU$ and $\pi_\cV$ are sub-Gaussian, we will prove that with arbitrarily high probability - exponential in $N$ -, there are only a polynomial number of such equations
    \item All those differential equations are then ``aggregated'' to build the final ODE satisfied by the matching size. Interestingly, this aggregated ODE has a simple form, while the full system is on the other hand quite intricate.
\end{enumerate}

In the following sub-sections, we separate the proofs in the different building blocks to provide intuitions; the proof of technical lemmas are deferred to the appendix.

\subsection{Building the graph together with the matching}

The first step in the analysis is to notice that the bipartite configuration model can be constructed by sequentially pairing the half-edges coming from $\cV$. The matching generated by \greedy\ is then constructed simultaneously with the graph. More precisely, given two sequences\footnote{Without loss of generality, we  assume that the additional extra vertex is always on the $\cV$ side.} of non-negative integers $\mathbf{d}^{\cU} = (d_1^{\cU}, \ldots, d_N^{\cU})$ and $\mathbf{d}^\cV \cup \{ d^{\cV}_{T+1} \} =(d_1^{\cV}, \ldots, d_T^{\cV}, d^{\cV}_{T+1})$, we introduce in the following a generating algorithm that simultaneously build the associated bipartite configuration model $\mathbf{CM}(\mathbf{d}^{\cU},\mathbf{d}^{\cV})$ together with \greedy. Recall that the bipartite configuration model is obtained through a uniform matching between the half-edges of $U$ and the half-edges of $V$. In order to avoid confusion, we will call a {\it marked matching} a pairing of two half-edges that corresponds to an edge that will belong to the constructed matching $\mathrm{M}$. This construction pseudo-code is detailed in Algorithm \ref{Algo:ConfGreedy}.

\begin{algorithm}
\SetAlgoLined
\DontPrintSemicolon
\KwIn{ $\mathbf{d}^\cU = (d_1^\cU, \ldots, d_N^\cU)$ and $\mathbf{d}^\cV=(d_1^\cV, \ldots, d_T^\cV)$}
 {\bf Initialization.}  $\mathrm{M}_0 \leftarrow \emptyset$, $\mathcal{E}_0\leftarrow \emptyset$ and $H^{\cU}_0 \leftarrow \{ \text{  half-edges  of } \cU\}$\;
 \For{$t= 1, \ldots, T$}{
  Order uniformly at random the edges emanating from $v_t$: $e_1^t, \ldots, e_{k_t}^t$ \;
  \For{$i = 1, \ldots, k_t$}{
  Choose uniformly an half-edge $e^\cU_i$ in $H^\cU$\;
   $\mathcal{E}\leftarrow \mathcal{E} \cup \{ u(e^\cU_i), v_t  \}$\tcp*{Create an edge between $e_i^t$ and $e^\cU_i$}
  $H^\cU\leftarrow H^\cU\setminus \{ e^\cU_i\}$\tcp*{Remove the half-edge}
  \If{$v_t$ and $u(e^\cU_i)$ unmatched}{
   $\mathrm{M}_t \leftarrow \mathrm{M}_{t-1} \cup \{ u(e^\cU_i), v_t  \}$\tcp*{$v_t$ is matched}
   }
  }
  }
 $\mathbf{CM}(\mathbf{d}^\cU,\mathbf{d}^\cV) \leftarrow (\cU,\cV,\cE)$.\\
 \KwOut{Bipartite configuration model $\mathbf{CM}(\mathbf{d}^\cU,\mathbf{d}^\cV)$ and matching $\mathrm{M}_T$ on it.}
 \caption{ \textsc{\greedy\ matching configuration model without capacities}  }
 \label{Algo:ConfGreedy}
\end{algorithm} 

Since each pairing of each half-edge is done uniformly at random, the graph obtained at the end of the algorithm has indeed the law of a bipartite configuration model. Moreover, it is easy to see that $\mathrm{M}$ corresponds to the matching constructed by \textsc{greedy matching} on $\mathbf{CM}(\mathbf{d}^\cU,\mathbf{d}^\cV)$.


\subsection{Differential Equation Method - Stochastic Approximation}
As mentioned above, several quantities are going to be tracked through time: for all $k \in \{0, \ldots, T\}$ and all $i\geq 0$, we define:
\begin{itemize}
\vspace{-0.5em}
\item $F_i(k)$ as the number of vertices  $u \in \cU$ that are not yet matched at the end of step $k$ and whose remaining degree is $i$, meaning that $d_u-i$ of their initial half-edges have been paired. We will  refer to them as  {\it free} vertices.\vspace{-0.5em}
\item $M_i(k)$ as the number of vertices $u \in \cU$  already matched at the end of step $k$ and whose remaining degree is $i$. We will  refer to them as {\it marked} vertices.
\end{itemize}
Notice that for all $0 \leq k \leq T$, the sum $F_i(k)+M_i(k)$ corresponds to the total number of vertices of $\cU$ with remaining degree $i$ at the end of step $k$. We also define
\begin{itemize}
\vspace{-0.5em}
\item $\widehat{F}(k):= \sum_{i \geq 0} i F_i(k)$ is the number of available half-edges attached to  free vertices at the end of step $k$,
\vspace{-0.5em}
\item $\widehat{M}(k) := \sum_{i \geq 0} i M_i(k)$ is the number of available half-edges attached to  marked vertices at the end of step $k$.
\end{itemize}
We are going to study the evolution of these quantities along with the one of \greedy. A major ingredient of the proof is to show that $F_i(k)$ and $M_i(k)$ closely follow the solutions of some ODE. This is the so-called {\it differential equation method} \citep{Wormald},  stated in Appendix \ref{Appendix_Wormald}.  For instance, it can easily be seen that $\widehat{F}(k)+\widehat{M}(k)$ closely follows the function $t\mapsto \mu_\cU-t\mu_\cV$ on $(0,\mu_\cU/\mu_\cV)$ in the following sense.

\begin{lemma}\label{lemma:Cvg_(F+M_hat)}
 For every $\varepsilon >0$, and for all $0 \leq k \leq T$,
\begin{equation*}
\bigg|\frac{\widehat{F}(k) + \widehat{M}(k)}{N} - \big(\mu_\cU- \frac{k}{N}\mu_\cV\big) \Big)\bigg| \leq \varepsilon.
\end{equation*}
with probability at least $1-\exp\big(-\frac{N\epsilon^2}{2\sigma_U^2}\big) + \exp\big(-\frac{T\epsilon^2}{2\sigma_V^2}\big)$.
\end{lemma}

We now turn to each individual quantity $F_i$ (resp. $M_i$). We can prove a similar result, yet the limit function is not explicit (unlike for the matching size as in Theorem \ref{theo:main_theo} statement). The following Lemma \ref{theo:FluidLimit_FetM} states that the discrete sequences of (free and marked) half-edges are closely related to the  solutions of some system of differential equations.

Before stating it, we first introduce, for any sequence of non-negative numbers $(x_\ell)_{\ell \geq 0}$ and $(y_\ell)_{\ell \geq 0}$ such that $0<\sum_\ell \ell (x_\ell +y_\ell)<\infty$, every $i \geq 0$, the following mappings
\begin{equation}\label{eq:Wormald:F_i}
\Phi_i(x_0,x_1, \ldots, y_0, y_1, \ldots ): = \frac{-i \mu_\cV x_i + (i+1) \mu_\cV x_{i+1} - h \left( \frac{\sum_{\ell\geq 0}\ell y_\ell}{\sum_{\ell\geq 0}\ell(x_\ell+y_\ell)} \right) (i+1) x_{i+1}}{\sum_{\ell \geq 0} \ell (x_\ell + y_\ell) }
\end{equation}
and
\begin{equation*}
\Psi_i(x_0,x_1, \ldots, y_0, y_1, \ldots ) := \frac{-i \mu_\cV y_i + (i+1) \mu_\cV y_{i+1} + h \left(\frac{\sum_{\ell\geq 0} \ell y_\ell}{\sum_{\ell\geq 0}\ell(x_\ell+y_\ell)} \right) (i+1) x_{i+1}}{\sum_{\ell\geq 0} \ell (x_\ell + y_\ell) },
\end{equation*}
 where $h$ is the following function, well-defined on $[0,1]$,
\[  h(s) = \frac{1 - \phi_\cV(s)}{1 - s}.  \]

\begin{lemma}\label{theo:FluidLimit_FetM}
With probability $1- \zeta N\exp(-\xi N^{c/2})$, there are at most $N^c$ quantities $F_i$ and $M_i$, and for all $0 \leq k \leq T$ and all $i \geq 0$
\begin{equation*}
\bigg|\frac{F_i(k)}{N} - f_i\left( \frac{k}{N} \right) \bigg| \leq \kappa N^{-2c} \quad \text{ and } \quad 
\bigg|\frac{M_i(k)}{N} - m_i \left( \frac{k}{N}  \right) \bigg| \leq  \kappa N^{-2c},
\end{equation*}
where $\zeta,\kappa$ depend only on the (first two) moments of $\pi_\cV$ and $\pi_\cU$ and $c=1/20$.

The continuous mappings $f_i$ and $m_i$ are solutions of the system of differential equations on $[0, \mu_\cU / \mu_\cV)$
\begin{equation}\label{EQ:system}
\begin{array}{ccl}
   \frac{\mathrm{d}f_i}{\mathrm{d}t} & = & \Phi_i(f_0, f_1, \ldots, m_0, m_1, \ldots),\\
   \frac{\mathrm{d}m_i}{\mathrm{d}t} & = & \Psi_i(f_0, f_1, \ldots, m_0, m_1, \ldots),\\
   f_i(0) & = & \pi_\cU(i),\\
   m_i(0) & = & 0.
\end{array}
\end{equation}
 \end{lemma}

This system is well defined as stated by the following Lemma \ref{lemma:uniqsol}.

\begin{lemma}\label{lemma:uniqsol}
The  system \eqref{EQ:system}  has a unique solution
which is well-defined on $[0, \mu_\cU / \mu_\cV)$. More precisely, denoting by $f$ and $m$ the generating series of the sequences $(f_i)_{i \geq 0}$ and $(m_i)_{i \geq 0}$,
\[  f(t,s) = \sum\limits_{i \geq 0} f_i(t) s^i \quad \quad \text{and} \quad \quad m(t,s)=\sum\limits_{i \geq 0} m_i(t) s^i,  \]
it holds that:
\begin{equation}\label{EQ:sol}
f\left(  \frac{\mu_\cU}{\mu_\cV} \left(1 - \e^{-\mu_\cV t} \right)  , s \right) = \phi_\cU \left( (s-1)e^{- \mu_\cV t} +1- F(t) \right),
\end{equation}
and
\begin{equation*}
 m\left(  \frac{\mu_\cU}{\mu_\cV} \left(1 - \e^{-\mu_\cV t} \right)  , s \right) = \int_{0}^{t} F'(u)\phi_\cU' \left( (s-1)e^{- \mu_\cV u} + 1 - F(u)  \right) \mathrm{d} u.
\end{equation*}
where $F$ is a solution of the following ODE \begin{equation*}
\frac{ \frac{1}{\mu_\cU} \phi_\cU'(1-F(t))}{1 - \phi_\cV \left( 1 - \frac{1}{\mu_\cU} \phi'_\cU(1-F(t))  \right)} F'(t) = \e^{- \mu_\cV t}.\end{equation*}
\end{lemma}

\subsection{Aggregating solutions to compute \greedy\ performances}

In order to get Theorem \ref{theo:main_theo}, notice that the number of vertices matched by \greedy\ is $N$ minus the number of free vertices remaining at the end, which is approximately equal to $Nf(\frac{\mu_\cU}{\mu_\cV},1)$ by definition of $f$ and because of Lemma \ref{theo:FluidLimit_FetM}. This corresponds to $t=+\infty$ in Equation \eqref{EQ:sol}, thus the performance of \greedy\ is, with arbitrarily high probability, arbitrarily close to 
$$
N(1-\phi_\cU(1-F(+\infty)))
$$
The statement of Theorem \ref{theo:main_theo} just follows from a simple final change of variable.

\section*{Conclusion}
We studied theoretical performances of \greedy\ algorithm on different matching problems with underlying structure. Those precise results are quite interesting and raise many questions, especially since \greedy\ actually outperforms \ranking\ in many different situation (in theory for $2$-regular graphs, but empirical evidences indicate that this happen more generically).

Our approach has also successfully  been used to unveil some  questions on the comparison between different possible models. But more general questions are still open; for instance, assuming that the expected degree is fixed, which situation is the more favorable to \greedy\ and online algorithm: small or high variance, or more generally this distribution $\pi_\cU$ or an alternative one $\pi_\cU'$ ? The obvious technique would be to compare the solution of the different associated ODE's. Similarly, the questions of stability/robustness of the solution to variation in the distribution $\pi_\cU$ and $\pi_\cV$ are quite challenging and left for future work.

\section*{Acknowledgments and Disclosure of Funding}

 V. Perchet acknowledges support from the ANR under grant number \#ANR-19-CE23-0026 as well as the support grant as part of the Investissement d’avenir project, reference ANR-11-LABX-0056-LMH, LabEx LMH, in a joint call with Gaspard Monge Program for optimization,operations research and their interactions with data sciences. Nathan Noiry also acknowledges support from the  Telecom Paris DSAIDIS chair.

\bibliographystyle{plainnat}
\bibliography{biblio}

\appendix

\section{General version of the result}\label{ap:generalresult}

\subsection{The fixed capacity matching problem}
\label{SE:Weights}
We now investigate the case where vertices $u \in \cU$ have capacities, which means that they can be matched to several vertices $v\in \cV$. Precisely, if the capacity of $u$ is denoted by $\omega_u$, then this vertex can be matched to at most $\omega_u$ vertices in $\cV$ (but as before, half-edges of $u$ are  going to be paired with $d_u$ half-edges originating from $\cV$). The graph is still constructed using the configuration model introduced in Section \ref{SE:ConfigModel}, i.e., the law of $d_u$ is $\pi_\cU$ (and similarly,  degrees of $v\in\cV$ are i.i.d., with law $\pi_\cV$).

For the moment, to simplify the analysis and the results statements, we are going to assume that all vertices $u \in \cU$ have the same initial capacity $C \in \mathbb{N}$. We denote the random graph with capacities generated this way by $\mathbf{CM}(\mathbf{d}^{\cU}, \mathbf{d}^{\cV},C)$

\begin{theorem}\label{theo:Fixed_Capacities}\textbf{(Performances of \greedy\  with fixed capacities)} 

Given $N \geq 1$ and $T= \frac{\mu_\cU}{\mu_\cV} N$, let $\mathrm{M}_T$ be the matching built by \greedy\ on $\mathbf{CM}(\mathbf{d}^{\cU}, \mathbf{d}^{\cV},C)$ then the following  convergence in probability holds:
\begin{equation*}\label{eq:convergence_probabilityCapCons}
\frac{|\mathrm{M}_T|}{CN} \overset{ \mathbf{P}}{ \underset{ N \rightarrow +\infty}{ \longrightarrow}} 1-\sum_{k=0}^{C-1}\frac{1-k/C}{k!}G(1)^{k}\phi_\cU^{(k)}\left(1-G(1)\right).
\end{equation*}
where $G$ is the unique solution of the following ordinary differential equation
\begin{equation*}
G'(s) = \frac{1 - \phi_\cV \left( 1 - \frac{1}{\mu_\cU}\Gamma_U(G(s))  \right)}{ \frac{\mu_\cV}{\mu_\cU}\Gamma_U(G(s))} .
\end{equation*} 
where
$$
\Gamma_U(g) = \phi'_\cU(1-g)+\sum_{k=1}^{C-1}\frac{g^k}{k!}\phi_\cU^{(k+1)}(1-g)
$$

Moreover, for any $s \in [0,1]$, if $M_{T}(s)$ is the matching obtained by \greedy\ after seeing a proportion  $s$ of vertices of $\cV$, then 
$$\frac{|\mathrm{M}_T(s)|}{CN} \overset{ \mathbf{P}}{ \underset{ N \rightarrow +\infty}{ \longrightarrow}} 1-\sum_{k=0}^{C-1}\frac{1-k/C}{k!}G(s)^{k}\phi_\cU^{(k)}\left(1-G(s)\right).$$
\end{theorem}

The proof of Theorem \ref{theo:Fixed_Capacities}, in Appendix \ref{Appendix_Fixed_Capacities}, has three major differences with the one of Theorem \ref{theo:main_theo}:

\begin{enumerate}
\item The first one is that more quantities must be tracked, not just the number of vertices with remaining free half-edges, but the number of such vertices for each possible values of remaining capacity;  the total number of equations is roughly speaking multiplied by a factor $(C+1)/2$ (since only $F_i(k)$ are affected by the capacities and not $M_i(k)$). We will therefore denote in the remaining by $F_i^{(c)}(k)$ the number of vertices with $i$ remaining half-edges to be paired and with current capacity equal to $c$ (those vertices can still be matched to $c$ different vertices $v \in \cV$).
 \item The second major difference lies in the resolution of the system of differential equations. The solution was rather direct without capacities (i.e., $c=1$). Unfortunately, the evolution of $F_{i}^{(c)}$ strongly depends on $F_{i}^{(c+1)}$. As a consequence, the trick is to solve this system by induction, starting from $c=C$ (this solution is almost identical to that of the case with no capacities) and then to inject this solution in the PDEs defining $F_i^{(C-1)}$ so on so forth. Indeed, the fluid limits  of $\sum_i F_i{(c)}$ and $\sum_i M_i$, that we denote respectively be $f^{(c)}$ and $m$ satisfy the following coupled equations (up to some time change  $\theta(t)$ and where $H(t)=h(q(t))$ for some function $q(\cdot)$ introduced in the proof):
\begin{equation*} 
\partial_t f^{(c)}(\theta(t),s) = \left[ - \mu_\cV s + \mu_\cV - H(t)\right] \partial_s f^{(c)}(\theta(t) ,s)+ H(t) \partial_s f^{(c+1)}(\theta(t),s)  ,
\end{equation*}
and
\begin{equation*} 
\partial_t m(\theta(t),s) = \left[- \mu_\cV s + \mu_\cV \right] \partial_s m(\theta(t),s) + H(t) \partial_s f^{(1)}(\theta(t),s).
\end{equation*}
\item Finally, the third main difference is how the performances of \greedy\ are defined. The upper-bound is obviously to create the minimum between $CN$ and $T$ matches (where $T$ is  the number of vertices in $\cV$). Anyway, those two numbers are within a constant multiplicative factor (recall that $T=\frac{\mu_\cU}{\mu_\cV}N$ for a valid configuration model), hence we arbitrarily chose to normalize \greedy\ performances by $CN$. As a consequence, the (normalized) performances of \greedy\ now rewrite as
$$
\frac{\sum_{i\geq0}\left(M_i(T)+\sum_{c=1}^C(1-\frac{c}{C}) F_i^{(c)}(T)\right)}{N},
$$
where $M_i(k)$ still denotes the number of marked vertices, i.e., those whose capacities have been depleted before step $k$ with $i$ remaining half-edges to be paired.
\end{enumerate}

\subsection{General case, online matching with capacities}
\label{SE:General}
In the general case, we no longer assume that all vertices $u \in \cU$ have the same initial capacities, but $\omega_u$ can be equal to any value in $\mathds{N}$ (yet this capacity is independent of the degree). Notice however that the capacities $\omega_u$ of vertices could be capped at their  degrees $d_u$ (since they would never be depleted otherwise). As a consequence,  capacities can be assumed to be bounded by $C< N^\beta$ for some $\beta <1$ since the maximal degree is also smaller than $N^\beta$ with arbitrarily high probability.

We therefore denote by $p_c \in [0,1]$ the fraction of vertices of $\cU$ whose initial capacity is exactly $c \in [1,C]$. Notice, we do not need to assume that capacities are drawn i.i.d.\ accordingly to some distribution, our results hold for any values $(p_c)_c$.  We denote by $\mathbf{CM}(\mathbf{d}^{\cU}, \mathbf{d}^{\cV},\mathbf{p})$ the random graph with capacities generated. 

Quite interestingly, the techniques are literally exactly the same as in  the previous case: we consider the exact same system of differential equations; the only differences are the initial conditions. Similarly, the maximal matching size is no longer $NC$ but $N\mathds{E}_{\mathbf{p}}[c] :=N\sum_c cp_c$. We also denote the cdf of the empirical distribution $p_c$ by $P_c:=\sum_{k\leq c} p_c$

\begin{theorem}\label{theo:General_Capacities}\textbf{(Performances of \greedy\  with different capacities)} 

Given $N \geq 1$ and $T= \frac{\mu_\cU}{\mu_\cV} N$, let $\mathrm{M}_T$ be the matching built by \greedy\ on $\mathbf{CM}(\mathbf{d}^{U}, \mathbf{d}^{V},\mathbf{p})$ then the following  convergence holds in probability:
\begin{equation*}\label{eq:convergence_probabilityCapVar}
\frac{|\mathrm{M}_T|}{N\mathds{E}_{\mathbf{p}}[c]} \overset{ \mathbf{P}}{ \underset{ N \rightarrow +\infty}{ \longrightarrow}} 1-\sum_{k=0}^{C-1}\frac{\sum_{c=1}^Ccp_{c+k}}{\mathds{E}_{\mathbf{p}}[c]}\frac{1}{k!}G(1)^{k}\phi^{(k)}\left(1-G(1)\right).
\end{equation*}
where $G$ is the unique solution of the following ordinary differential equation
\begin{equation*}
G'(s) = \frac{1 - \phi_\cV \left( 1 - \frac{1}{\mu_\cU}\Gamma^{\mathbf{p}}_U(G(s))  \right)}{ \frac{\mu_\cV}{\mu_\cU}\Gamma^{\mathbf{p}}_U(G(s))} .
\end{equation*} 
with
$$
\Gamma^{\mathbf{p}}_U(g)) = \phi'_U(1-g)+\sum_{k=1}^{C-1}\left(\frac{(1-P_k)g^k}{k!}\phi_\cU^{(k+1)}(1-g)\right).
$$
Moreover, for any $s \in [0,1]$, if $M_{T}(s)$ is the matching obtained by \greedy\ after seeing a proportion  $s$ of vertices of $\cV$, then 
$$\frac{|\mathrm{M}_T(s)|}{N\mathds{E}_{\mathbf{p}}[c]} \overset{ \mathbf{P}}{ \underset{ N \rightarrow +\infty}{ \longrightarrow}} 1-\sum_{k=0}^{C-1}\frac{\sum_{c=1}^Ccp_{c+k}}{\mathds{E}_{\mathbf{p}}[c]}\frac{1}{k!}G(s)^{k}\phi^{(k)}\left(1-G(s)\right).$$
\end{theorem}
As mentioned before, the proof (delayed to Appendix \ref{Appendix:Proof_General_Capacities}) is rather similar to the previous one; the major difference is that the change of initial condition of the system of PDE makes it a bit more complicated to solve (hence the more intricate formulation of the result).

\section{Additional Numerical Experiments}
\subsection{Further comparisons between the theoretical result and simulations}

\begin{figure}[ht!]
\includegraphics[width=0.33\textwidth]{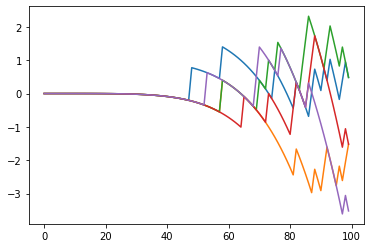}
\includegraphics[width=0.33\textwidth]{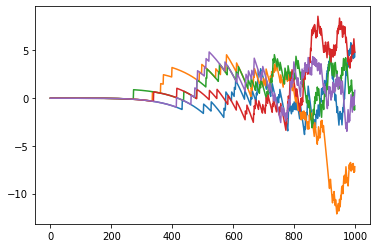}
\includegraphics[width=0.33\textwidth]{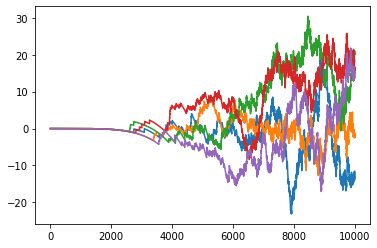}
\caption{Difference between the theoretical performances and simulated performances of the \greedy\ algorithm on the $d$-regular graph ($d=4$) on 5 independent runs, with $N=100,1000,10000$.}
\label{FIG:dregular}
\end{figure}

\begin{figure}[ht!]
\includegraphics[width=0.33\textwidth]{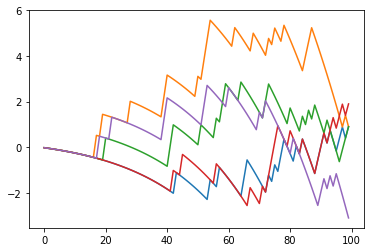}
\includegraphics[width=0.33\textwidth]{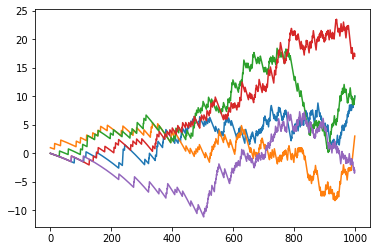}
\includegraphics[width=0.33\textwidth]{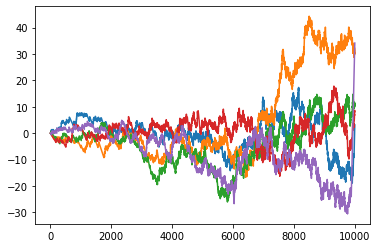}
\caption{Difference between the theoretical value \ref{eq:ODE_on_F_atS} and simulated performances of the \greedy\ algorithm on the Erd\H{o}s-R\'enyi graph, $c=4$, on 5 independent runs, with $N=100,1000,10000$.}
\label{FIG:erdos}
\end{figure}

We provide in Figure \ref{FIG:dregular} a comparison between the score predicted by the numerical solutions of the ODE (the functions $1-\phi_\cU(1-G(s))$) for $4$-regular graphs and the simulated performance of \greedy\ for various values of $N$. As expected, the deviations of the simulated trajectories remain within $\mathcal{O}(\sqrt{N})$ of the expected theoretical trajectory. Figure \ref{FIG:erdos} illustrates the same comparison on an Erd\H{o}s-R\'enyi graph whose expected degree equals $4$.

\begin{figure}[ht!]
\centering
\includegraphics[scale=0.4]{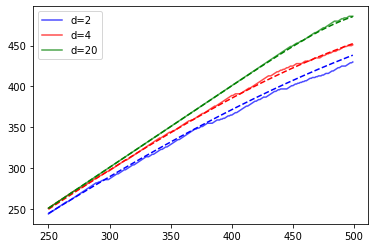}
\includegraphics[scale=0.4]{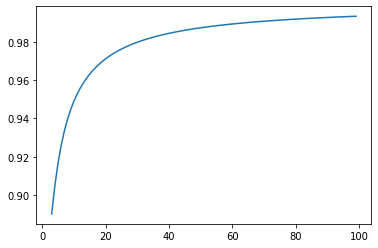}
\caption{On the left, expected theoretical performance of the \greedy\ algorithm (dashed line) along with the simulated performance (full line) for various values of $d$. On the right, expected competitive ratio of \greedy\ on the $d$-regular graph as a function of $d$.}
\label{FIG:dregularscore2}
\end{figure}

In Figure \ref{FIG:dregularscore2}, we plot the theoretical performance of the \greedy\ algorithm along with its experimental performance on the $d$-regular graph for various values of $d$. We also  plot the competitive ratio of \greedy\ predicted by the ODE as function of $d$. As expected, the score increases with $d$ (as the problem becomes simpler and simpler for \greedy\, when $d\geq 2$).

\subsection{\greedy\ vs \ranking}
We illustrate in this section the quite surprising fact that, in some configuration model, \greedy\ actually outperforms \ranking. We recall that the latter algorithm chooses at random a ranking over $\cU$ and uses it to break ties (i.e., if two vertices $u$ and $u'$ can be matched to $v_k$, then it is the one with the smallest rank that is matched by \ranking).

In adversarial configuration, it is known that the competitive ratio of \ranking\ is $1-\frac{1}{e}$ which is bigger than the one of \greedy, equal to $1/2$, see \citep{Mehta}. In the following figures, we also plot the performances of two other ``algorithms'' \smallest\ and \highest, for the sake of comparison; indeed, those are not admissible algorithms as they use the (future) knowledge of the number of half-edges of each vertex $u\in\cU$.

More precisely, \smallest\ matches a vertex $v_k \in \cV$ to the vertex $u \in \cU$ with the smallest number of remaining half-edges (under the constraints obviously that $(u,v_k) \in \cE$). As a consequence \smallest\ could be seen as an upper limit for online algorithm.

\highest\ does the opposite: it matches $v_k$ to the vertex $u \in \cU$ with the highest remaining number of half-edges. So \highest\ should serve as a lower bound/sanity check for any online algorithm.

\begin{figure}[ht!]
\begin{center}
\includegraphics[width=0.3\textwidth]{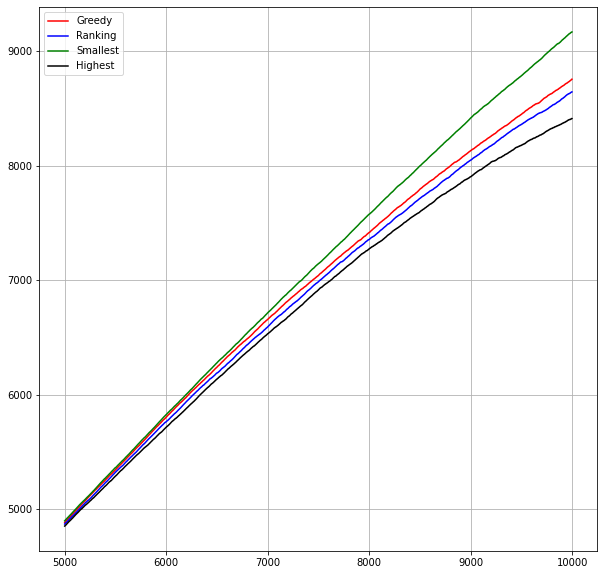}\includegraphics[width=0.3\textwidth]{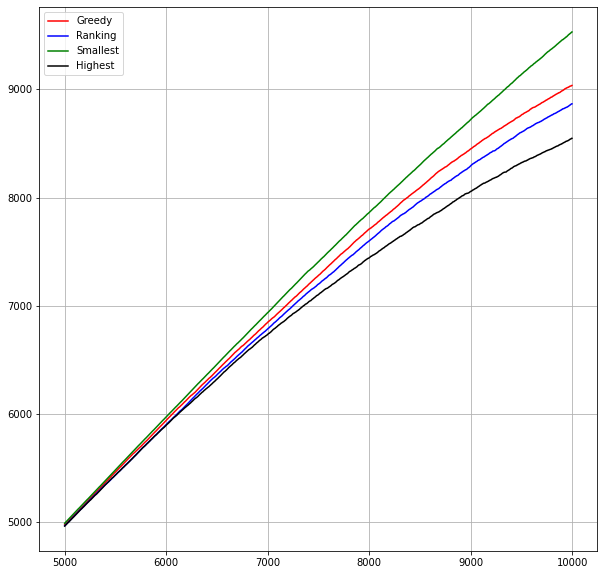}

\includegraphics[width=0.3\textwidth]{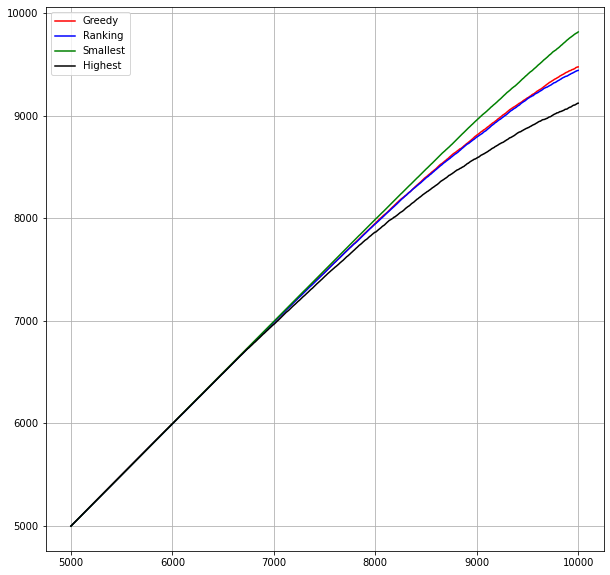}\includegraphics[width=0.3\textwidth]{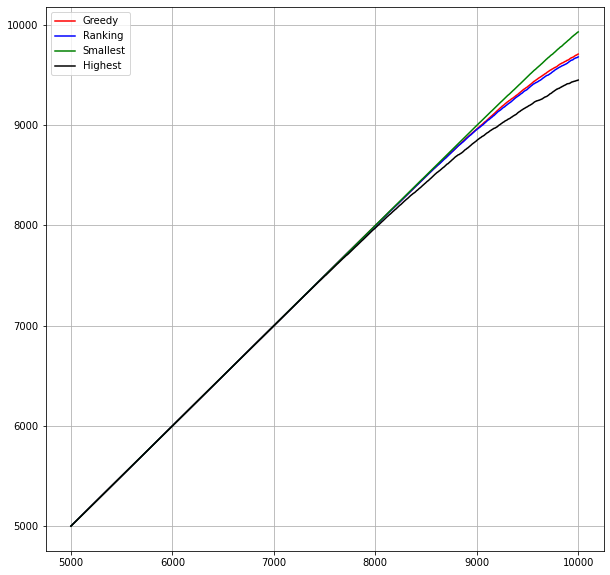}
\caption{\greedy\ outperforms \ranking\ in some configuration models }
\label{FIG:Fourmatchings}
\end{center}
\end{figure}

In Figure \ref{FIG:Fourmatchings}, the performances of those 4 matching ``algorithm'' (again \smallest\ and \highest\ are not admissible as they use extra knowledge) are illustrated on configuration models with $d=2, 4, 10$ and $20$.

As mentioned before, \greedy\ surprisingly outperforms \ranking\ in some configuration models, with a relative performance that decreases with $d$ (which is rather natural on the other hand, since the relative performance of \highest\ and \smallest\ also decreases). 

Figure \ref{FIG:Fourmatchings} also illustrates the different time steps at which algorithms fail to match new vertices $v_k$ (because all the $u$ they are paired with are already matched with another vertex $v_j$ for some $j<k$). This happens later and later as $d$ increases (as expected), at around half the horizon for $d=2$ and roughly $82\%$ with $d=20$.

\subsection{A few vertices with high capacity vs many vertices with low capacity}

In this section, we investigate how nodes capacities  affect \greedy's expected performance. The baseline is its performance on a random graph where all vertices have capacity $1$ and the vertices degrees in $\cU$ and $\cV$ follow the distributions $\pi_\cU$ and $\pi_\cV$. The comparison graph with capacity $C$ has $|\cU|/C$ "in-place" vertices, each with a capacity $C$, and their degrees follows the modified distribution  $\tilde{\pi}^U_C$ where $\tilde{\pi}^U_C(x=k)=\pi_\cU(x=k/C)$. Informally, the graph with capacity $C$ is built from the baseline graph by merging $C$ vertices of equal degree $d$ into a single vertex of degree $dC$.

\begin{figure}[ht!]
\centering
\includegraphics[width=0.3\textwidth]{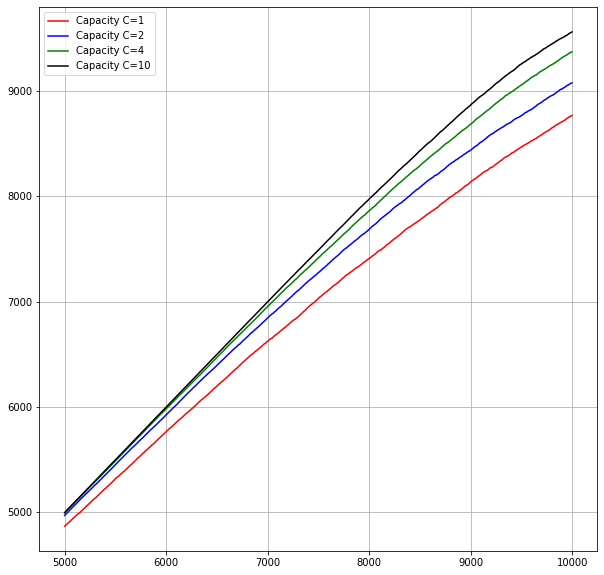}\includegraphics[width=0.3\textwidth]{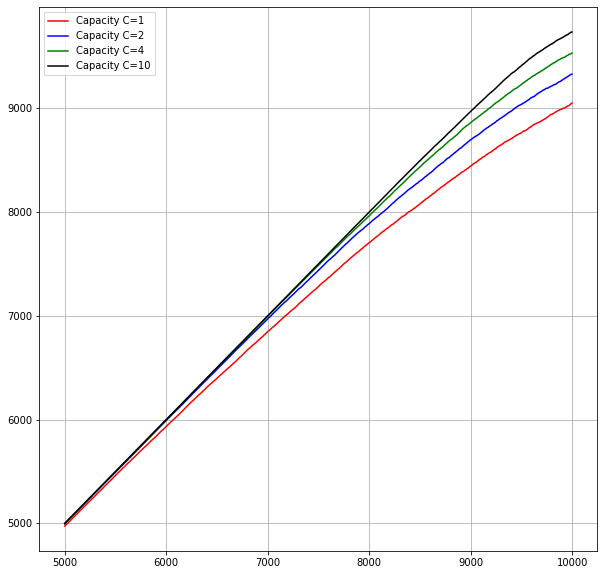}
\includegraphics[width=0.3\textwidth]{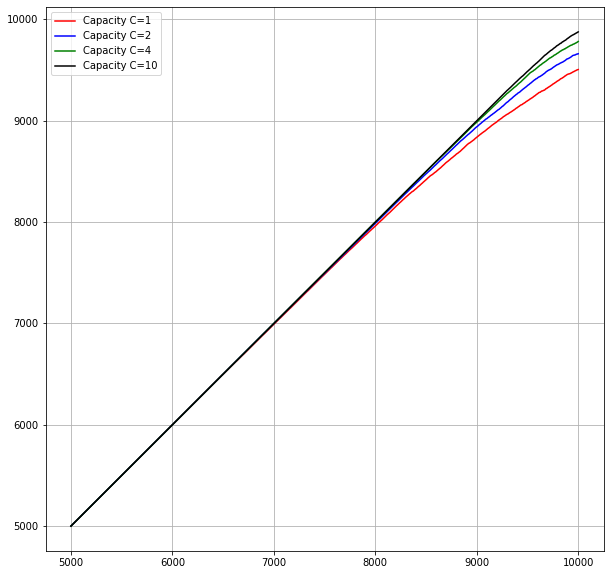}\includegraphics[width=0.3\textwidth]{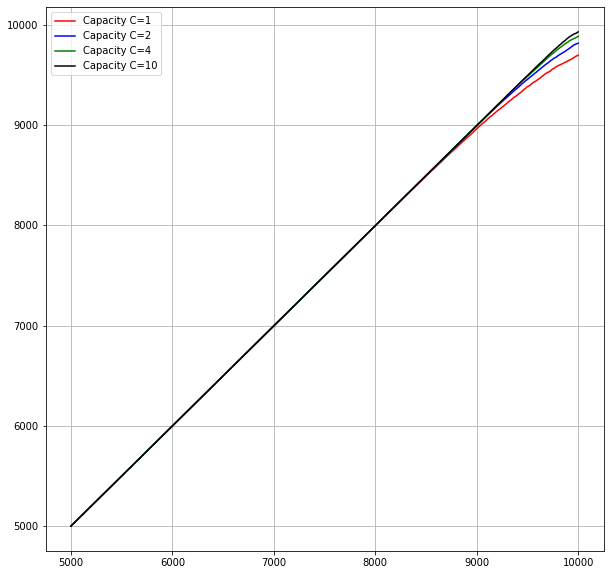}
\caption{\greedy\ performs better in high capacity graphs }
\label{FIG:Fourmatchings2}
\end{figure}

The results of the simulation illustrate that the \greedy\  performs better on graphs with vertices of high capacity.


\section{Stochastic approximation \& Differential equation method}
\label{Appendix_Wormald}
The following  theorem is an improved version of Wormald's Theorem \citep{Noiry}.
\begin{theorem}\label{theo:Wormald}
Let $a > 0$. For all $ N \geq 1$ and all $1 \leq k \leq N^a$, let $Y_k(i) =Y_k^{(N)}(i)$ be a Markov chain with respect to a filtration $\{\mathcal{F}_i\}_{i \geq 1}$. Suppose that, for all $k \geq 1$, there exists a function $f_k$ such that:
\begin{itemize}
\item $Y_k(0)/N = z_k(0)$;
\item $| Y_{k}(i+1) - Y_k(i) | \leq N^\beta $;
\item $ \big| \mathbf{E} \left[ Y_k(i+1) - Y_k(i) \, \Big| \, \mathcal{F}_i  \right] - f_k \left( \frac{i}{N}, \frac{(Y_k(i))_{1 \leq k \leq N^a}}{N}  \right) \big| \leq  c N^{-\lambda}$, for some constant $c>0$
\end{itemize}
where $0 < \beta < 1/2$, $\lambda >0$. Suppose that the following infinite system of differential equations with initial conditions $(z_k(0))_{k \geq 1}$ has a unique solution $(z_k)_{k \geq 1}$:
\[ \forall k \geq 1, \quad    z_k'(t) = f_k(t,(z_k(t))_{k \geq 1}). \]
Then, for all $k \geq 1$, $Y_k(\lfloor tN \rfloor)/N$ converges in probability towards $z_k$ for the topology of uniform convergence.

\medskip

More precisely, for every $1 < \varepsilon < \frac{1-\beta}{\beta}$, for every $\frac{(1+\varepsilon)\beta}{2} < \alpha < \varepsilon \beta $ and for every $0 \leq i\leq \frac{N}{\omega}$ where $\omega = N^{(1+\varepsilon)\beta}$, it holds that 
$$
\mathds{P}\Big(|Y(i\omega)-z(\frac{i\omega}{N})N| \leq i \big(N^{\alpha + \beta}+cN^{(1+\varepsilon)\beta-\lambda}+N^{2(1+\varepsilon)\beta-2}\big)\Big) \leq i \exp\Big(-\frac{N^{2\alpha-(1+\varepsilon)\beta}}{2}\Big)
$$
\end{theorem}

\section{Proofs of technical steps of Theorem \ref{theo:main_theo}} 
\label{Appendix_main_theo}
\subsection{Proof of Lemma \ref{lemma:Cvg_(F+M_hat)}}
It is an application of (maximal) Hoeffding-Azuma inequality  since, for every $0 \leq k \leq M-1$,
\[  \mathbf{E} \left[ \big(\widehat{F}(k+1) + \widehat{M}(k+1) \big) - \big(\widehat{F}(k) + \widehat{M}(k) \big) \, \big| \, \mathcal{F}_k \right] = - \mathbf{E}\left[ d_k^{\cV} \right] = - \mu_\cV.  \]

\subsection{Proof of Lemma \ref{theo:FluidLimit_FetM}}

Since $\pi_\cV$ is $\sigma_{\cV}$ subGaussian, then for any $\beta> 0$, 
\[  \mathbf{P}\left(  \exists i \in \{1, \ldots,T\}, \, d^\cV_i \geq \mu_\cV + N^\beta  \right)  \leq T \exp(-\frac{N^{2\beta}}{2\sigma^2_{\cV}}).  \] 
In particular, for some $\beta <1/2$ to be chosen later on, if $\mu_\cV \leq N^\beta/2$, then all degrees are smaller than $N^\beta$ with probability at least $1-T\exp\Big(-\frac{N^{2\beta}}{8\sigma_\cV^2}\Big)$; from now on, we will place ourselves on that event. 

\medskip


We also denote by $(\mathcal{F}_k)_{0 \leq k \leq M}$ the natural filtration associated to the \textsc{greedy matching} algorithm. In order to apply Theorem \ref{theo:Wormald}, it remains to control  for every $i \geq 0$ and  $0 \leq k \leq M-1$,
$$\bigg|\mathbf{E}\left[ F_i(k+1) - F_i(k) \, \big| \, \mathcal{F}_k   \right] - \Phi_i \left(  F_0 \left( \frac{k}{N}\right), F_1 \left( \frac{k}{N} \right), \ldots, M_0\left( \frac{k}{N} \right), M_1 \left( \frac{k}{N} \right), \ldots  \right)\bigg|
$$
and
$$\bigg|\mathbf{E}\left[ M_i(k+1) - M_i(k) \, \big| \, \mathcal{F}_k   \right] - \Psi_i \left(  F_0 \left( \frac{k}{N}\right), F_1 \left( \frac{k}{N} \right), \ldots, M_0\left( \frac{k}{N} \right), M_1 \left( \frac{k}{N} \right), \ldots  \right)\bigg| $$

 Let $0 \leq k \leq T- \frac{2N^\gamma}{\mu^\cV}$, with $\gamma >1/2$ some parameter to be fixed later,  so that, according to Lemma \ref{lemma:Cvg_(F+M_hat)}, with probability at least $1-\exp(-\frac{N^{2\gamma-1}}{2\sigma_{\cU}^2})-\exp(-\frac{\mu_\cU N^{2\gamma-1}}{2\mu_\cV\sigma_{\cV}^2})$  it holds that $\widehat{F}(k) + \widehat{M}(k) \geq N^\gamma$.

 Recall that, in the $k$-th step of the algorithm,  half-edges of the $k$-th vertex of $V_N$ are ordered uniformly  at random: $(e_i^k)_i$ for $i=1, \ldots, d^\cV_k$. Then,  each of these half-edges is sequentially paired uniformly at random with half-edges of $d^\cV$ that are not yet paired. Let $u_i^k$ be the vertex to which $e_i^k$ is paired and let $I_k$  be the first integer $i$ such that $u_i^k$ belongs to the free vertices of $\cU$ at time $k$, that is to the vertices that are not yet matched. If such an integer does dot exist, that is when all $u_i^k$ are already matched, we set $I_k= +\infty$.  As a consequence, we aim at estimating $\mathbf{P} \left( I_k =i \, \big| \, \mathcal{F}_k  \right)$ for the different admissible values, where this probability has the following explicit definition
\begin{align*}
\mathbf{P} \left( I_k =i \, \big| \, \mathcal{F}_k  \right)
&= \frac{\widehat{M}(k)}{\widehat{F}(k) + \widehat{M}(k)} \frac{\widehat{M}(k)-1}{\widehat{F}(k)+\widehat{M}(k)-1} \cdots \frac{\widehat{M}(k)-(i-2)}{\widehat{F}(k)+\widehat{M}(k)-(i-2)} \frac{\widehat{F}(k)}{\widehat{F}(k)+\widehat{M}(k)-(i-1)}\\
&= \frac{\widehat{M}(k) !}{(\widehat{M}(k)-(i-1))!}\frac{(\widehat{F}(k)+\widehat{M}(k)-i)!}{(\widehat{F}(k)+\widehat{M}(k))!}\widehat{F}(k)
\end{align*}
First, assume that $\widehat{M}(k) \geq 2N^\theta$ for some parameter $\theta > 2\beta$ to be chosen later, so that those probabilities are all strictly positive. Using Stirling approximation formula, we get that, with $p(k)= \frac{\widehat{M}(k)}{\widehat{F}(k) + \widehat{M}(k)}$ and for any $i$,
$$
0 \geq \frac{\mathbf{P} \left( I_k =i \, \big| \, \mathcal{F}_k  \right) - (1-p(k))^{i-1}p(k)}{(1-p(k))^{i-1}p(k)}  \geq  -2\frac{N^{2\beta}}{N^\theta} - \frac{N^\beta}{N^\gamma} 
$$
Second, assume that $\widehat{M}(k) < 2N^\theta$ for some $\theta > \beta$. This immediately implies that, for $i$,
$$0 \geq \mathbf{P} \left( I_k =i \, \big| \, \mathcal{F}_k  \right) - (1-p(k))^{i-1}p(k) \geq -2\frac{N^\theta}{N^\gamma}$$ 
Similar inequalities holds  for $\mathbf{P}( I_k = +\infty \, \big| \, \mathcal{F}_k)$, except that it is approximately equal to $\mathbf{E}\left[ (1-p(k))^{d_k^\cV} \right]=\phi_\cV(1-p(k))$.

\medskip

It remains to control the evolution of the processes $F_i(k)$ and $M_i(k)$. Notice that, by their very definition, on  the event $I_k = x$ for some $1 \leq x \leq d_k^\cV$, the following happens:
\begin{enumerate}
\item The first $x-1$ half-edges $e_k^1, \ldots, e_k^x$ are paired uniformly at random with marked half-edges of $\cU$. If the corresponding vertex has remaining degree equal to $i$, then $M_i$ decreases by one and $M_{i-1}$ increases by one.

\item The $x$-th half-edge $e_k^x$ is paired uniformly at random with free half-edge of $\cU$. If the corresponding vertex has remaining degree $i$, then $F_i$ decreases by one and $M_{i-1}$ increases by one.

\item The $d_k^\cV-x$ remaining half-edges $e_k^{x+1}, \ldots, e_k^{d_k^\cV}$ are paired uniformly at random with half-edges of $\cU$. If the corresponding vertex is free with remaining degree $i$, then $F_i$ decreases by one and $F_{i-1}$ increases by one. Otherwise, if the corresponding vertex is marked with remaining degree $i$, then $M_i$ decreases by one and $M_{i-1}$ increases by one.
\end{enumerate}
Notice that, after the pairing of each half-edges, the quantity $\widehat{F}(k)$ (resp. $\widehat{M}(k)$) may decrease (resp. increase) by one. Therefore, working on the event where $d_k^\cV \leq N^\beta$, we deduce that $\widehat{F}$ and $\widehat{M}$ are affected by an additive term of order at most $N^\beta$. The same argument holds on $F_i$ and $M_i$. 

All of these considerations imply that
\begin{align*}
& \bigg| \mathbf{E} [  F_i(k+1) - F_i(k) \, \big| \, \mathcal{F}_k, \, I_t = x ] -\Big(- \frac{iF_i(k)}{\widehat{F}(k)} + ( \mu_\cV -x) \left( - \frac{iF_i(k)}{\widehat{F}(k)+\widehat{M}(k)} + \frac{(i+1)F_{i+1}(k)}{\widehat{F}(k)+\widehat{M}(k)}   \right) \Big)  \bigg| \\ &\leq  2\sigma^2_{\mu_\cV}\frac{N^\beta}{N^\gamma} \end{align*}
and similarly
\begin{align*}
\Bigg|\mathbf{E} [ M_i(k+1) - M_i(k)  \, \big| \, \mathcal{F}_k, \, I_t=x ] -  \bigg(&(x-1) \Big( - \frac{iM_i}{\widehat{M}} + \frac{(i+1)M_{i+1}}{\widehat{M}}   \Big)
+ \frac{(i+1)F_{i+1}}{\widehat{F}}\bigg)  \\
&  + ( \mu_\cV -x) \left( - \frac{iM_i}{\widehat{F}+\widehat{M}} + \frac{(i+1)M_{i+1}}{\widehat{F}+\widehat{M}}   \right) \bigg) \Bigg|  \\
\leq& 2\sigma^2_\mu\frac{N^\beta}{N^\gamma}+2 \sum_{j=1}^{x-1}\frac{j}{\widehat{M}(k)-j}
\end{align*}

Finally, the case $I_t = +\infty$ is handled similarly, as by definition
\begin{equation*}
 \mathbf{E} [ F_i(k+1) -F_i(k)  \, | \, \mathcal{F}_k, \, I_t = +\infty ] = 0.
\end{equation*}
and the following also holds also holds:
$$ \bigg| \mathbf{E} [ M_i(k+1) - M_i(k) \, | \, \mathcal{F}_k, \, I_t = +\infty ] -  \mu_\cV  \left( - \frac{iM_i}{\widehat{M}} + \frac{(i+1)M_{i+1}}{\widehat{M}} \right) \bigg| \leq 2\sigma^2_\mu\frac{N^\beta}{N^\gamma}+2 \sum_{j=1}^{d_k^\cV-1}\frac{j}{\widehat{M}(k)-j}.$$

It remains to compute the expected variation in $F_i(k)$ and $M_i(k)$. It is a bit simpler for the former, but still, to  to lighten the notations, we write $p=p_k$ and $q=q_k$ in the following computation.

\begin{align*}
 \mathbf{E}&\big[ F_i(k+1) - F_i(k) \, \big| \, \mathcal{F}_k   \big] \\
&= \mathbf{E}_{d_k^\cV \sim \pi_\cV} \left[ \sum\limits_{x=1}^{d_k^\cV} q^{x-1} \left(  - \frac{i F_i}{\widehat{F} + \widehat{M}} \right) + p \sum\limits_{x=1}^{d_k^\cV} q^{x-1} (d_k^\cV-x) \left( - \frac{i F_i}{\widehat{F} + \widehat{M}} + \frac{(i+1) F_{i+1}}{\widehat{F} + \widehat{M}} \right) \right] + \eta_N \\
&= \frac{1}{\widehat{F} + \widehat{M}} \mathbf{E}_{d_k^\cV \sim \pi_\cV} \left[  -iF_i \left( \sum_{i=1}^{d_k^\cV} q^{x-1} + p d_k^\cV \sum\limits_{x=1}^{d_k^\cV} q^{x-1} - p \sum\limits_{x=1}^{d_k^\cV} x q^{x-1}  \right) \right.  \\ 
& \hspace{5cm} \left. + (i+1)F_{i+1} \left( p d_k^\cV \sum\limits_{x=1}^{d_k^\cV} q^{x-1} - p \sum\limits_{x=1}^{d_k^\cV} x q^{x-1}  \right)   \right] + \eta_N\\
&= \frac{1}{\widehat{F} + \widehat{M}} \mathbf{E}_{d_k^\cV \sim \pi_\cV} \Bigg[ -iF_i \left( \frac{1-q^{d_k^\cV}}{1-q} + d_k^\cV (1-q^{d_k^\cV}) - \frac{d_k^\cV q^{d_k^\cV +1} - (d_k^\cV+1)q^{d_k^\cV} +1}{1-q}  \right) \\
&\hspace{5cm} + (i+1)F_{i+1} \left( q (1-q^{d_k^\cV}) - \frac{d q^{d_k^\cV+1} - (d_k^\cV+1)q^{d_k^\cV} + 1}{1-q}  \right) \Bigg] + \eta_N \\
&= \frac{1}{\widehat{F} + \widehat{M}} \mathbf{E}_{d_k^\cV \sim \pi_\cV} \left[ -iF_i d_k^\cV + (i+1) F_{i+1} \frac{q^{d_k^\cV} - d_k^\cV q + (d_k^\cV-1)}{1-q}  \right] +\eta_N \\
&= \frac{1}{\widehat{F} + \widehat{M}} \mathbf{E}_{d_k^\cV \sim \pi_\cV} \left[ -iF_i +(i+1) d_k^\cV F_{i+1} + (i+1) F_{i+1} \frac{1 - q^{d_k^\cV}}{1-q}  \right] + \eta_N \\
&= \frac{-i\mu_\cV F_i + (i+1) \mu_\cV F_{i+1} - (i+1) h(q) F_{i+1} }{\widehat{F} + \widehat{M}} + \eta_N,
\end{align*}
which is exactly \eqref{eq:Wormald:F_i},  up to error term $\eta_N$ that satisfies, if $\widehat{M}(k) < 2 N^\theta$,
$$
|\eta_N | \leq 2\sigma^2_{\mu_\cV}\frac{N^\beta}{N^\gamma} + 2\mu^2_\cV \frac{N^\theta}{N^\gamma}
$$ and, if $\widehat{M}(k) \geq 2 N^\theta$,
$$
|\eta_N| \leq 3\sigma^2_{\mu_\cV}\frac{N^\beta}{N^\gamma} + 2\mu_\cV^2 \frac{N^{2\beta}}{N^\theta} + \mu_\cV^2\frac{N^\beta}{N^\gamma}
$$

Computations are quite similar for the difference in $M_i(k)$ and the error term  still depends whether $\widehat{M}(k)$ is bigger, or smaller, than $2N^\theta$:
\begin{align*}
    \mathbf{E}&\big[ M_i(k+1) - M_i(k) \, \big| \, \mathcal{F}_k   \big] \\
    &= \mathbf{E}_{d_k^\cV \sim \pi_\cV} \left[ \left( \frac{d_k^\cV(1-p)^{d_k^\cV}}{\widehat{M}} +\sum\limits_{x=1}^{d_k^\cV}pq^{x-1}\left(\frac{(x-1)}{\widehat{M}}+\frac{d_k^\cV-x}{\widehat{M}+\widehat{F}}\right)\right)\Big((i+1)M_{i+1}-iM_i\Big)\right]\\
    &\hspace{2cm}+\mathbf{E}_{d_k^\cV \sim \pi_\cV} \left[\sum\limits_{x=1}^{d_k^\cV}pq^{x-1}\frac{(i+1)F_{i+1}}{\widehat{F}}\right] + \varepsilon_N  \\
    &= \frac{1}{\widehat{M}+\widehat{F}}\mathbf{E}_{d_k^\cV \sim \pi_\cV} \left[ \left( d_k^\cV(1-p)^{d_k^\cV-1} +\sum\limits_{x=1}^{d_k^\cV}\left(pq^{x-2}(x-1)+pq^{x-1}(d_k^\cV-x)\right)\right)\Big((i+1)M_{i+1}-iM_i\Big)\right]\\
    &\hspace{2cm}+\frac{(i+1)h(q)F_{i+1}}{\widehat{M}+\widehat{F}} +\varepsilon_N  \\
    &= \frac{\mu_\cV((i+1) M_{i+1}-iM_i) +(i+1) h(q) F_{i+1} }{\widehat{F} + \widehat{M}} +  \varepsilon_N,
\end{align*}
where   $\varepsilon_N$ satisfies, if $\widehat{M}(k) < 2N^\theta$, 
$$
|\varepsilon_n| \leq 2\sigma^2_\cV \frac{N^\theta}{N^\gamma}+2\sigma^2_\cV \frac{N^\beta}{N^\gamma}+2\mu^2_\cV \frac{N^\theta}{N^\gamma};
$$
and, if $\widehat{M}(k) \geq 2N^\theta$, it satisfies 
$$
|\varepsilon_n|\leq 3\sigma^2_\cV \frac{N^\beta}{N^\gamma}+2\mu_\cV^2 \frac{N^{2\beta}}{N^\theta} + \mu_\cV^2\frac{N^\beta}{N^\gamma};
$$
We used in the above computations (at  the third equality)  the following observation:
\begin{align*}
    kq^{k-1}+\sum_{x=1}^kpq^{x-1}\left(\frac{x-1}{q}+k-x\right)=k
\end{align*}

Summing error terms over all the $2N^\beta$ equations relating $F_i$ to $f_i$ and $M_j$ to $m_j$, the error terms coming from the differential equation method Theorem \ref{theo:Wormald}, and using the fact that $m$ is $\mu_\cU$-Lipschitz, we get that the total error, defined by,
$$\mathrm{Err}:=\sup_{s \in [0,1]}\Big| \frac{|\mathrm{M}_T(s)|}{N} - \big(1 - \phi_\cU(1-G(s))\big) \Big|$$
satisfies 
$$\mathrm{Err} \leq N^\beta N^{(1+\varepsilon)\beta}(N^{\alpha +\beta}+4(\sigma^2_\cV+\mu^2_\cV)N^{(1+\varepsilon)\beta -\gamma/2+\beta}+N^{2(1+\varepsilon)\beta-2})+\mu_\cU N^{(1+\varepsilon)\beta}$$
as soon as $\theta=\beta + \frac{\gamma}{2}$.

It remains to pick admissible values for the different parameters, such as the following ones (checking admissibility follows from immediate computations):
$$ \beta = 1/20, \epsilon = 10, \gamma = 21/40, \theta = 25/80, \alpha. 23/80$$
Those choices ensures that $\mathrm{Err} = \mathcal{O}(N^{-1/20})$. 

All those arguments hold with probability at least (summing all the bad event probabilities)
$$
1-T\exp(-\frac{N^{2\beta}}{2\sigma_V^2}) - \exp(-\frac{N^{2\gamma -1}}{2\sigma^2_\cU})- \exp(-\frac{\mu_\cU N^{2\gamma -1}}{2\mu_\cV\sigma^2_\cV}) - 2N^\beta N^{(1+\varepsilon)\beta)}\exp(-N^{2\alpha - (1+\varepsilon)\beta})  \geq 1 - \mathcal{O}( N\exp(-\zeta N^{1/40}))
$$
where the equality holds because of the choice of parameters.

\subsection{Proof of Lemma \ref{lemma:uniqsol}}

Notice that the functions $f$ and $m$ satisfy the following partial differential equations:
\begin{equation*} \label{equ:PDE_f}
\partial_t f(t,s) = \frac{1}{\mu_\cU - t \mu_\cV} \left[ - \mu_\cV s + \mu_\cV - h(q(t))   \right] \partial_s f(t,s),
\end{equation*}
and
\begin{equation*} \label{equ:PDE_m}
\partial_t m(t,s) = \frac{1}{\mu_\cU - t \mu_\cV} \left[ - \mu_\cV s + \mu_\cV \right] \partial_s m(t,s) + h(q(t)) \partial_s f(t,s),
\end{equation*}
where $q(t)= \partial_s f(t,1)/(\mu^\cU-t\mu^\cV)$.

In order to solve these equations, we first perform a time change to get rid of the denominator. Let 
\begin{equation*}
\theta(t) = \frac{\mu_\cU}{\mu_\cV} \left( 1 - \e^{ - \mu_\cV t}  \right)
\end{equation*}
so that $\theta'(t) = \mu_\cU - \theta(t) \mu_\cV$. In order to simplify notations, we set:
\begin{equation*}
H(t) := h \left(q( \theta(t)) \right).
\end{equation*}
 
Then, the new functions
\[ g(t,s) := f(\theta(t),s) \quad \quad \text{and} \quad \quad o(t,s) := m (\theta(t),s) \]
satisfy the following PDEs:
\begin{equation} \label{eq:EDP_1}
\partial_t g(t,s) =  \left[ - \mu_\cV s + \mu_\cV - H(t)   \right] \partial_s g(t,s),
\end{equation}
and
\begin{equation} \label{eq:EDP_2}
\partial_t o(t,s) = \left[ - \mu_\cV s + \mu_\cV \right] \partial_s o(t,s) + H(t) \partial_s g(t,s).
\end{equation}
These two equations fall into the classical framework of {\it transport} differential equation and can be explicitly solved. We give the details for the reader's convenience.

\paragraph{Solution of \eqref{eq:EDP_1}.}
Let $s$ be a solution of the following ODE:
\begin{equation} \label{eq:ODE_1}
s'(t) = \mu_\cV s(t) - \mu_\cV + H(t).
\end{equation}
Then, the function $g$ is constant along the curve $(t,s(t))$. Indeed:
\[ \frac{\mathrm{d}}{\mathrm{d}t}g(t,s(t)) = \partial_t g(t,s) + s'(t) \partial_s g(t,s) = 0.   \]
The differential equation \eqref{eq:ODE_1} admits the following general solutions:
\[  s_c(t) = \left[ c + e^{-\mu_\cV t} - 1 + \int_0^t e^{- \mu_\cV u} H(u) \mathrm{d}u   \right] e^{ \mu_\cV t}.  \]
Therefore,
\[  (t,s) = (t,s_c(t)) \, \,  \Longleftrightarrow  \, \, c = c(t,s) = (s-1)e^{- \mu_\cV t} + 1 - \int_0^t e^{- \mu_\cV u} H(u) \mathrm{d}u,  \]
and we deduce that (the initial condition is $g(0,s) = \phi_\cU(s)$):
\begin{equation}  \label{eq:Solution_g}
g(t,s) = g(0,c(t,s)) = \phi_\cU(c(t,s)) = \phi_\cU \left( (s-1)e^{- \mu_\cV t} + 1 - \int_0^t e^{- \mu_\cV u} H(u) \mathrm{d}u  \right).  
\end{equation}

\paragraph{Solution of \eqref{eq:EDP_2}.}
Let $s_\gamma(t) = \gamma e^{ \mu_\cV t} + 1$. Then, $s'_\gamma(t) = \mu_\cV s(t) - \mu_\cV$ and we deduce that, along the curves $(t,s_\gamma(t))$, $o(t,s)$ satisfies the following ODE:
\[ \frac{\mathrm{d}}{\mathrm{d}t}o(t,s_\gamma(t)) = \frac{1-q(t)^d}{1-q(t)} \partial_s g(t,s_{\gamma}(t)). \]
Since
\[ (t,s) = (t,s_\gamma(t)) \, \, \Longleftrightarrow \, \, \gamma = \gamma(t,s) = (s-1)e^{- \mu_\cV t}, \]
we deduce that:
\begin{equation}\label{eq:Solution_o}
o(t , s)=\int_{0}^{t} H(u) \partial_{s} g(u, (s-1)e^{-\mu_\cV (t-u)}+1) \mathrm{d} u. 
\end{equation}

We now define the function $F(\cdot)$ as
\begin{equation}
F(t) := \int_0^t \e^{ - \mu_\cV u} H(u) \mathrm{d}u.
\end{equation}
Using Equations \eqref{eq:Solution_g} and \eqref{eq:Solution_o}, one can easily deduce that
\[  \partial_s g(t,1) = \e^{-\mu_\cV t} \phi_\cU'( 1 - F(t))  \]
and
\begin{align*}  
\partial_s o(t,1) &= \int_0^t H(u) \e^{- \mu_\cV u} \phi''_U( 1 - F(u)) \mathrm{d}u  \\ 
&= \phi'_U(1) - \phi'_U(1-F(t)) = \left(\mu_\cU  - \phi_\cU'(1-F(t)) \right) \e^{- \mu_\cV t}.
\end{align*}
In particular,
\[ \partial_s g(t,1) + \partial_s o(t,1) = \partial_s f(\theta(t),1) + \partial_s m(\theta(t),1) = \mu_\cU \e^{- \mu_\cV t}.   \]
Therefore,

\begin{align*}
H(t) = \frac{1 - \phi_\cV\left( \frac{\partial_s o(t,1)}{\partial_s g(t,1) +\partial_s o(t,1) } \right)}{1 - \frac{\partial_s o(t,1)}{\partial_s g(t,1) +\partial_s o(t,1)}}  = \mu_\cU \frac{1 - \phi_\cV \left( 1 - \frac{1}{\mu_\cU \phi_\cU'(1 - F(t))}   \right)}{1 - \phi_\cU( 1 -F(t) )},
\end{align*}

which yields the following ordinary differential equation on $F$:
\begin{equation}
\frac{ \frac{1}{\mu_\cU} \phi_\cU'(1-F(t))}{1 - \phi_\cV \left( 1 - \frac{1}{\mu_\cU} \phi'_U(1-F(t))  \right)} F'(t) = \e^{- \mu_\cV t}.
\end{equation}

\section{Proof of Theorem \ref{theo:Fixed_Capacities}}
\label{Appendix_Fixed_Capacities}

We recall the notations introduced. For all $k \in \{0, \ldots, T\}$, all $c \in \{0, \ldots, C\}$ and all $i\geq 0$, we define:
\begin{itemize}\item $F_i^{(c)}(k)$ the number of vertices of $\cU$ that still have capacity $c$ at the end of step $k$ and whose remaining degree is $i$. Those vertices are referred to as {\it free} (with remaining degree $i$ and capacity $c$ at the end of step $k$). 

\item $M_i(k)$ the number of vertices of $\cU$ that have capacity $c=0$ at the end of step $k$ and whose remaining degree is $i$. Those vertices are referred to as {\it marked} (with remaining degree $i$ at the end of step $k$).
\end{itemize}
We also define as before the number of remaining half-edges to respectively free and marked vertices as
$$
\widehat{F}(k) = \sum_{c=1}^C\sum_{i\geq 0}i F_i^{(c)}(k),\quad \text{ and } \quad  \widehat{M}(k) = \sum_{i\geq 0}i M_i(k).
$$

The normalized performances of \greedy\ is the ratio between the matched vertices in $\cV$ and its maximal number, equal to $CN$:
$$
A = \frac{\sum_{i\geq0}\left(CM_i(T)+\sum_{c=1}^C(C-c) F_i^{(c)}(T)\right)}{CN}
$$

As in the proof of Theorem \ref{theo:main_theo}:
\begin{enumerate}
    \item we will place ourselves on the event where all vertices have bounded degrees, smaller than $N^\beta$ for some small $\beta>0$
    \item we will stop the analysis at $N^\gamma$ steps of the horizon $T$ so that $\widehat{F}(k)+\widehat{M}(k) > N^\gamma$ with arbitrarily high probability 
    \item  we will distinguish the cases where $\widehat{M}(k) > 2 N^\theta$ (with $\theta = \beta +\gamma/2$)
\end{enumerate}
As a consequence, the errors are going to be of the same order of magnitude with the same order of probability (up to a multiplicative factor $C)$ (hence those computations are skipped and replace by $\mathcal{O}(\cdot)$ notations). The interesting new component in this proof is the new system of differential equations and their solutions.
\subsection{The Differential equations}
Using the same notations than in the proof of Theorem \ref{theo:main_theo}, we  get that for all $0 \leq k \leq T$, $i \geq 0$ and $c \leq C$,
\begin{align*}
 \mathbf{E}&\big[ F_i^{(c)}(k+1) - F_i^{(c)}(k) \, \big| \, \mathcal{F}_k  \big] \\
&= \mathbf{E}_{d_k \sim \pi_\cV} \left[ \sum\limits_{x=1}^{d_k} -q^{x-1}\frac{i F_i^{(c)}}{\widehat{F} + \widehat{M}}  + p \sum\limits_{x=1}^{d_k} q^{x-1} (d_k-x) \left(\frac{(i+1) F_{i+1}^{(c)}-i F_i^{(c)}}{\widehat{F} + \widehat{M}} \right) \right]  \\
&\hspace{5cm}+\mathbf{E}_{d_k \sim \pi_\cV} \left[ \sum\limits_{x=1}^{d_k} q^{x-1} \frac{(i+1)F_{i+1}^{(c+1)}}{\widehat{F} + \widehat{M}} \right]+ \mathcal{O}(N^{\theta-\gamma})\\
&= \frac{\mu_\cV\left(-i F_i^{(c)} + (i+1) F_{i+1}^{(c)} \right)- (i+1) h(q) F_{i+1}^{(c)}+(i+1)h(q)F_{i+1}^{(c+1)} }{\widehat{F} + \widehat{M}} + \mathcal{O}(N^{\theta-\gamma})
\end{align*}
where the function $h$ is still defined as $h(q) = \frac{1-\phi_\cV(q)}{1-q}$. Similarly, we can compute the expected increment in $M_i$ as
\begin{align*}
    \mathbf{E}&\big[ M_i(k+1) - M_i(k) \, \big| \, \mathcal{F}_k   \big] \\
    &= \mathbf{E}_{d_k \sim \pi_\cV} \left[ \left( \frac{d_k(1-p)^{d_k}}{\widehat{M}} +\sum\limits_{x=1}^{d_k}pq^{x-1}\left(\frac{(x-1)}{\widehat{M}}+\frac{d_k-x}{\widehat{M}+\widehat{F}}\right)\right)\left((i+1)M_{i+1}-iM_i\right)\right]\\
    &\hspace{2cm}+\mathbf{E}_{d_k \sim \pi_\cV} \left[\sum\limits_{x=1}^{d_k}pq^{x-1}\frac{(i+1)F_{i+1}^{(1)}}{\widehat{F}}\right] + \mathcal{O}(N^{\theta-\gamma}) \\
    &= \frac{\mu_\cV((i+1) M_{i+1}-iM_i) +(i+1) h(q) F^{(1)}_{i+1} }{\widehat{F} + \widehat{M}} + \mathcal{O}(N^{\theta-\gamma})
\end{align*}

From this, we get the following system of differential equations:

\begin{equation} \label{equ:PDE_fbis}
\partial_t f^{(c)}(t,s) = \frac{1}{\mu_\cU - t \mu_\cV} \left[\left( - \mu_\cV s + \mu_\cV - h(q(t))   \right) \partial_s f^{(c)}(t,s)+\frac{1}{\mu_\cU - t \mu_\cV} h(q(t)) \partial_s f^{(c+1)}(t,s) \right] ,
\end{equation}
and
\begin{equation} \label{equ:PDE_mbis}
\partial_t m(t,s) = \frac{1}{\mu_\cU - t \mu_\cV} \left[\left( - \mu_\cV s + \mu_\cV \right) \partial_s m(t,s) + h(q(t)) \partial_s f^{(1)}(t,s)\right] 
\end{equation}

\bigskip
With those notations, the normalized performances of \greedy\ rewrite then  into:
\begin{align*}
 A &= m(\frac{\mu_\cU}{\mu_\cV},1)+\sum_{c=1}^C(1-\frac{c}{C})f^{(c)}(\frac{\mu_\cU}{\mu_\cV},1)  \\
 &= 1-\sum_{c=1}^C\frac{c}{C}f^{(c)}(\frac{\mu_\cU}{\mu_\cV},1)
\end{align*}

\subsection{Solving the PDEs}

As in the previous section, we start with a time change. Let 
\begin{equation}
\theta(t) = \frac{\mu_\cU}{\mu_\cV} \left( 1 - \e^{ - \mu_\cV t}  \right)
\end{equation}
so that $\theta'(t) = \mu_\cU - \theta(t) \mu_\cV$. In order to simplify notations, we set:
\begin{equation}
H(t) := h \left(q( \theta(t)) \right).
\end{equation}
Then, the new functions
\[ g^{(c)}(t,s) := f^{(c)}(\theta(t),s) \quad \quad \text{and} \quad \quad o(t,s) := m (\theta(t),s) \]
satisfy the following PDEs:
\begin{equation*} \label{eq:EDP_1c}
\partial_t g^{(c)}(t,s) =  \left[ - \mu_\cV s + \mu_\cV - H(t)   \right] \partial_s g^{(c)}(t,s)+H(t)\partial_sg^{(c+1)}(t,s),
\end{equation*}
and
\begin{equation} \label{eq:EDP_2c}
\partial_t o(t,s) = \left[ - \mu_\cV s + \mu_\cV \right] \partial_s o(t,s) + H(t) \partial_s g^{(1)}(t,s).
\end{equation}

We distinguish:
\begin{equation} \label{eq:EDP_3}
\partial_t g^{(C)}(t,s) =  \left[ - \mu_\cV s + \mu_\cV - H(t)   \right] \partial_s g^{(C)}(t,s)
\end{equation}

We define:
$$
F(t) = \int_0^t e^{- \mu_\cV u} H(u) \mathrm{d}u 
$$
\paragraph{Solution of \eqref{eq:EDP_3}.}

This equation is the same as the one satisfied by $g(t,s)$, with the same initial conditions. Thus, we can write:

\begin{equation*}  \label{eq:Solution_g2}
g^{(C)}(t,s) = \phi_\cU \left( (s-1)e^{- \mu_\cV t} + 1 - F(t)  \right).  
\end{equation*}

\paragraph{Solution of \eqref{eq:EDP_3}.}
Lets define the curves:
\[  s_{t,s}(u) = \left[ (s-1)e^{- \mu_\cV t} - F(t)  + F(u)   \right] e^{ \mu_\cV u}+1.  \]

Along those curves, we have:
\[ 
\frac{\mathrm{d}}{\mathrm{d}t}g^{(c)}(u,s_{t,s}(u)) = H(u)\partial_s g^{(c+1)}(u,s_{t,s}(u)). 
\]

So:
\[
g^{(c)}(t,s) = \int_0^t H(u) \partial_s g^{(c+1)}(u,s_{t,s}(u)) \mathrm{d}u
\]

\subparagraph{Solution for $c=C-1$.}
We have:
\begin{align*}
    g^{(c-1)}(t,s) &= \int_0^t H(u) \partial_s g^{(C)}(u,s_{t,s}(u)) \mathrm{d}u\\
  &= \int_0^t F'(u) \phi'_U((s-1)e^{- \mu_\cV u}+1-F(u))\mathrm{d}u\\
  &=  F(t)\phi'_U((s-1)e^{- \mu_\cV t}+1-F(t))
\end{align*}

\subparagraph{Solution for $c=C-k$, general formula.}
We will prove by induction:
\[
g^{(C-k)}(t,s) = \frac{1}{k!}(F(t))^k\phi^{(k)}((s-1)e^{- \mu_\cV t}+1-F(t))
\]
If it is true for rank $k$, we have:
$$
\partial_s g^{(C-k)}(u,s_{t,s}(u))= \frac{e^{- \mu_\cV u}}{k!}(F(u))^k\phi^{(k+1)}\left((s-1)e^{- \mu_\cV t}+1-F(t)\right)
$$
Which gives:
\begin{align*}
    g^{(C-(k+1))}(t,s) &= \frac{1}{k!}\left(\int_0^tF'(u)(F(u))^k \mathrm{d}u\right)\phi^{(k+1)}\left((s-1)e^{- \mu_\cV t}+1-F(t)\right)\\
  &= \frac{1}{(k+1)!}(F(t))^{k+1}\phi^{(k+1)}((s-1)e^{- \mu_\cV t}+1-F(t))
\end{align*}

\paragraph{Solution of \eqref{eq:EDP_2c}.}

Let's define the curves:
\[
\gamma_{s,t}(u) = 1 + (s-1)e^{-\mu_\cV(t-u)}
\]
Along those curves:
\[
\frac{\mathrm{d}}{\mathrm{d}u}o(u,\gamma_{t,s}(u))= H(u)\partial_s g^{(1)}(u,\gamma_{t,s}(u))
\]

So:
\begin{align*}
    o(t,s) = \int_0^t F'(u) \frac{(F(u))^{(C-1)}}{(C-1)!}\phi^{(C)}\left((s-1)e^{- \mu_\cV t}+1-F(u)\right)\mathrm{d}u
\end{align*}

\paragraph{Formula for \greedy\ performances.}
Recall that the normalized performances of \greedy\ are 
\begin{align*}
 A &= 1-\sum_{c=1}^C\frac{c}{C}g^{(c)}(+\infty,1)\\
 &= 1-\sum_{k=0}^{C-1}\frac{1-\frac{k}{C}}{k!}(F(+\infty))^{k}\phi^{(k)}\left(1-F(+\infty)\right)\\
\end{align*}

\subsection{ODE for F}

We have as before:
$$
F'(t) = H(t)e^{- \mu_\cV t}, \ H(t) = \frac{1-\phi_\cV(Q(t))}{1-Q(t)}
$$
And we also have:
$$
Q(t) =\frac{\partial_s o(t,1)}{\partial_s o(t,1)+\sum_{c=1}^C \partial_s g^{(c)}(t,1)}
$$

According to the previous section :
\begin{align*}
    \partial_s o(t,1) &= \left(\int_0^t F'(u) \frac{(F(u))^{(C-1)}}{(C-1)!}\phi_\cU^{(C+1)}\left(1-F(u)\right)\mathrm{d}u\right)e^{- \mu_\cV t}\\
    &= \left(\int_0^{F(t)}  \frac{x^{(C-1)}}{(C-1)!}\phi_\cU^{(C+1)}\left(1-x\right)\mathrm{d}x\right)e^{- \mu_\cV t}\\
    &= \left[\phi'_U(1)-\phi'_U(1-F(t))-\sum_{k=1}^{C-1}\frac{F(t)^K}{k!}\phi_\cU^{(k+1)}(1-F(t))\right]e^{- \mu_\cV t}
\end{align*}

Which gives:
$$
Q(t) = 1-\frac{1}{\mu_\cU}\left(\phi'_U(1-F(t))+\sum_{k=1}^{C-1}\frac{F(t)^k}{k!}\phi_\cU^{(k+1)}(1-F(t))\right)
$$

We define:
$$
\Gamma_U(F(t)) = \frac{1}{\mu_\cU}\left(\phi'_U(1-F(t))+\sum_{k=1}^{C-1}\frac{F(t)^k}{k!}\phi_\cU^{(k+1)}(1-F(t))\right)
$$
This yields the following differential equation for $F$:
\begin{equation*}
\frac{ \Gamma_U(F(t))}{1 - \phi_\cV \left( 1 - \Gamma_U(F(t))  \right)} F'(t) = \e^{- \mu_\cV t}.
\end{equation*}

Theorem \ref{theo:Fixed_Capacities} then follows from the same arguments in the proof of Theorem \ref{theo:main_theo} (except that errors are $C$
 times bigger as there are $C$ more equations to handle).

\section{Proof of Theorem \ref{theo:General_Capacities}}
\label{Appendix:Proof_General_Capacities}
As mentioned in the main text, the only differences with Theorem \ref{theo:Fixed_Capacities} is that $C$ could be  of the order of  $N^\beta$ (but not bigger on the event where all degrees are smaller than $N^\beta$). As a consequence, one must take $\beta$ even smaller than $1/20$ to have sublinear errors terms (choosing $\beta =1/40$ is  admissible for instance) with exponentially high probability.
\paragraph{Solution of \eqref{eq:EDP_3}.}

This equation is the same as the one satisfied by $g(t,s)$, the new initial condition is $g^{(C)}(t,s)= p_C \phi_\cU(s)$. Thus, we can write:

\begin{equation*} 
g^{(C)}(t,s) = p_C\phi_\cU \left( (s-1)e^{- \mu_\cV t} + 1 - F(t)  \right).  
\end{equation*}

\subparagraph{Solution for $c=C-1$.}

We have:
\begin{align*}
    g^{(c-1)}(t,s) &= \int_0^t H(u) \partial_s g^{(C)}(u,s_{t,s}(u)) \mathrm{d}u+ g^{(c-1)}(0,s_{t,s}(0))\\
  &= \int_0^t F'(u) \phi'_U((s-1)e^{- \mu_\cV t}+1-F(t))\mathrm{d}u+p_{(C-1)}\phi_\cU((s-1)e^{- \mu_\cV t} - F(t)+1)\\
  &=  p_CF(t)\phi'_U((s-1)e^{- \mu_\cV t}+1-F(t))+p_{(C-1)}\phi_\cU((s-1)e^{- \mu_\cV t} +1- F(t))
\end{align*}

\subparagraph{Solution for $c=C-2$.}
$$
\partial_s g^{(C-2)}(u,s_{t,s}(u))= p_Ce^{- \mu_\cV u}F(u)\phi_\cU''\left((s-1)e^{- \mu_\cV t}+1-F(t)\right)+p_{(C-1)}e^{- \mu_\cV u}\phi_\cU'(s_{t,s}(u))
$$

Let's define:
$$
c(t,s) = (s-1)e^{- \mu_\cV t}+1-F(t)
$$

Which gives:
\begin{align*}
    g^{(C-2)}(t,s) =& p_{(C-1)}\left(\int_0^tF'(u)F(u) \mathrm{d}u\right)\phi^{''}\left(c(t,s)\right)\\
    &+p_{(C-1)}\left(\int_0^tF'(u)e^{ \mu_\cV u}\mathrm{d}u\right)\phi^{'}\left(c(t,s)\right) \mathrm{d}u+p_{(C-2)}\phi_\cU(s_{t,s}(0))\\
  =& \frac{p_C}{2}(F(t))^{2}\phi^{''}((s-1)e^{- \mu_\cV t}\\
  &+1-F(t))+p_{(C-1)}F(t)\phi^{'}((s-1)e^{- \mu_\cV t}+p_{(C-2)}\phi_\cU(c(t,s))
\end{align*}

\subparagraph{Solution for $c=C-k$, general formula.}
We will prove by induction:
\[
g^{(C-k)}(t,s) = \sum_{l=0}^{k}p_{C-l}\frac{1}{(k-l)!}(F(t))^{k-l}\phi^{(k-l)}(c(t,s))
\]
If it is true for rank $k$, we have:
$$
\partial_s g^{(C-k)}(u,s_{t,s}(u))= \sum_{l=0}^{k}p_{C-l}\frac{1}{(k-l)!}(F(t))^{k-l}e^{- \mu_\cV u}\phi^{(k+1-l)}(c(t,s))
$$
Which gives:
\begin{align*}
    g^{(C-(k+1))}(t,s) &= p_{(C-(k+1))}\phi_\cU(c(t,s))+\sum_{l=0}^{k}p_{(C-l)}\frac{1}{(k-l)!}\left(\int_0^tF'(u)(F(u))^{(k-l)} \mathrm{d}u\right)\phi^{(k+1-l)}\left(c(t,s)\right)\\
  &=\sum_{l=0}^{k+1}p_{(C-l)}\frac{1}{(k+1-l)!}(F(t))^{k+1-l}\phi^{(k+1-l)}(c(t,s))
\end{align*}

\paragraph{Solution of \eqref{eq:EDP_2c}.}

\begin{align*}
    o(t,s) = \sum_{c=1}^{C} p_c\int_0^t F'(u) \frac{(F(u))^{(c-1)}}{(c-1)!}\phi^{(c)}\left((s-1)e^{- \mu_\cV t}+1-F(u)\right)\mathrm{d}u
\end{align*}

\[
g^{(c)}(t,s) = \sum_{k=0}^{C-c}p_{c+k}\frac{1}{k!}(F(t))^{k}\phi^{k}(c(t,s))
\]

\paragraph{Quantity of interest.}
\begin{align*}
 A &= \frac{\mu_\cV}{\mu_\cU}\sum_{c=1}^Cc(p_c-g^{(c)}(+\infty,1))\\
 &= \frac{\mu_\cV}{\mu_\cU}\left(\sum_{c=1}^Cc p_c-\sum_{k=0}^{C-1}\left(\frac{1}{k!}(F(+\infty))^{k}\phi^{(k)}\left(1-F(+\infty)\right)\sum_{c=1}^Ccp_{c+k}\right)\right)\\
\end{align*}

\paragraph{ODE for the function F.}

\begin{align*}
    \partial_s o(t,1) &= \left(\sum_{c=1}^Cp_c\int_0^t F'(u) \frac{(F(u))^{(c-1)}}{(c-1)!}\phi_\cU^{(c+1)}\left(1-F(u)\right)\mathrm{d}u\right)e^{- \mu_\cV t}\\
    &= \left[\phi'_U(1)-\phi'_U(1-F(t))+\sum_{k=1}^{C-1}\left(\frac{F(t)^k}{k!}\phi_\cU^{(k+1)}(1-F(t))\sum_{c=k+1}^Cp_c\right)\right]e^{- \mu_\cV t}
\end{align*}

Which yields:

$$
Q(t) = 1-\frac{1}{\mu_\cU}\left(\phi'_U(1-F(t))+\sum_{k=1}^{C-1}\left(\frac{F(t)^k}{k!}\phi_\cU^{(k+1)}(1-F(t))\sum_{c=k+1}^Cp_c\right)\right)
$$

We define:
$$
\Gamma_U(F(t)) = \frac{1}{\mu_\cU}\left(\phi'_U(1-F(t))+\sum_{k=1}^{C-1}\left(\frac{F(t)^k}{k!}\phi_\cU^{(k+1)}(1-F(t))\sum_{c=k+1}^Cp_c\right)\right)
$$
This yields the following differential equation for $F$:
\begin{equation*}
\frac{ \Gamma_U(F(t))}{1 - \phi_\cV \left( 1 - \Gamma_U(F(t))  \right)} F'(t) = \e^{- \mu_\cV t}.
\end{equation*}

\section{ Proof of \cref{prop:greedyvsranking}}\label{ap:greedyvsranking}

\begin{lemma}
On the $2$-regular graph, the law of the matches generated by the algorithm Ranking equals the law of the matches generated by a biased Greedy algorithm, that chooses a free vertex of degree $2$ over one of degree $1$ with probability $2/3$. This is biased as the classical Greedy algorithm chooses it with probability $1/2$.
\end{lemma}

\textit{Proof}: Two vertices of same degree are interchangeable, they are both equally likely to have the smallest rank. Thus Ranking and Greedy behave the same on an arriving vertices with potential neighbors of same degree. Let $r(v)$ be the rank of vertex $v$ and $deg(v)$ its residual number of unpaired half-edges.

\begin{align*}
    \PP(r(v)=k|\text{deg}(v)=1) =& \PP(\max(r(a),r(b))=k|\text{deg}(a)=\text{deg}(b)=2)\\
        =&2\frac{k-1}{n(n-1)}
\end{align*}

\begin{align*}
    \PP(r(v)<r(u)|\text{deg}(v)=2,\text{deg}(v)=1) =& \sum_{k=1}^n 2\frac{k-1}{n(n-1)}\cdot\frac{k-2}{n-2}\\
    =&\frac{2}{3}
\end{align*}

\hfill \(\Box\)

Let $M_1^G(t)$ and $M_1^R(t)$ be the number of marked vertices of degree $1$ by the \greedy\ and \ranking\ algorithms respectively. Note that the number of vertices of degree $2$ is the same for both algorithm, $F_2^G(t)=F_2^R(t)$. Also, the following always holds
$$F_1^G(t)=2N-2t-2F_2^G(t)-M_1^G(t).$$ 
Suppose it holds at time $t$ that $M_1^G(t)=M_1^R(t)=M_1(t)$ (event $\mathcal{A}$), then 
\begin{align*}
    \E[M_1^R(t+1)|\mathcal{A}]-\E[M_1^G(t+1)|\mathcal{A}] = &\E[\mathds{1}_{\{\ranking \text{ marks a vertex in }  F_2^R(t)\}}|\mathcal{A}]-\E[\mathds{1}_{\{\greedy \text{ marks a vertex in }  F_2^R(t)\}}|\mathcal{A}]\\
    &=\frac{1}{6}\cdot\frac{F_1(t)F_2(t)}{2(N-t)}>0\\
\end{align*}

Therefore, \ranking\ generates strictly more marked vertices of degree $1$ on average. As the probability that an incoming vertice is matched only to non-avalaible vertices increases with $M_1$, \ranking\ performs stricly worse than \greedy\ on this $2$-regular graph.
\end{document}

%% file: prelude.tex
\usepackage[utf8]{inputenc}
\usepackage[english]{babel}

\usepackage{amsmath}
\usepackage{amsthm}
\usepackage{amssymb,bm}

\usepackage{amssymb,amsmath,amsthm, amsfonts}

\usepackage{natbib}

\newcommand{\cV}{\mathcal{V}}
\newcommand{\cU}{\mathcal{U}}
\newcommand{\cE}{\mathcal{E}}
\newcommand{\cM}{\mathcal{M}}

\newcommand{\N}{\mathds{N}}

\newcommand{\E}{\mathds{E}}

\usepackage[utf8]{inputenc}
\usepackage[english]{babel}
\usepackage{amsfonts}
\usepackage{bm}
\usepackage{color}
\usepackage{mathrsfs}  
\usepackage[expansion]{microtype}
\usepackage{esint}
\usepackage{graphicx}
\usepackage{dsfont}
\usepackage[dvipsnames]{xcolor}         
\usepackage[]{hyperref}
\usepackage[capitalise, noabbrev]{cleveref}

\newtheorem{theorem}{Theorem}
\newtheorem{lemma}{Lemma}
\newtheorem{corollary}{Corollary}
\newtheorem{proposition}{Proposition}

\usepackage[ruled,vlined]{algorithm2e}

\usepackage{hyperref}
\usepackage{comment}

\newcommand{\greedy}{\textsc{greedy}}
\newcommand{\ranking}{\textsc{ranking}}
\newcommand{\smallest}{\textsc{smallest}}
\newcommand{\highest}{\textsc{highest}}

\newcommand\PP{ \mathbf{P} }

\usepackage[textwidth=30mm]{todonotes}

\DeclareMathOperator{\e}{e}